\documentclass[12pt]{iopart}

\expandafter\let\csname equation*\endcsname\relax
\expandafter\let\csname endequation*\endcsname\relax
\usepackage{amsmath}
\usepackage{amssymb}
\usepackage{hyperref}
\usepackage{graphicx}
\usepackage{mathrsfs}
\usepackage{fullpage}
\usepackage[usenames,dvipsnames]{color}
\usepackage{ulem}

\numberwithin{equation}{section}

\begin{document}

\title{Spectral Characteristic Evolution: A New Algorithm for Gravitational Wave Propagation}
\author{Casey J. Handmer, B\'{e}la Szil\'{a}gyi}
\address{Theoretical Astrophysics 350-17, California Institute of Technology, Pasadena, California 91125, USA}
\ead{chandmer@caltech.edu}
\begin{abstract}
We present a spectral algorithm for solving the full nonlinear vacuum Einstein field equations in the Bondi framework. Developed within the Spectral Einstein Code (\texttt{SpEC}), we demonstrate spectral characteristic evolution as a technical precursor to Cauchy Characteristic Extraction (CCE), a rigorous method for obtaining gauge-invariant gravitational waveforms from existing and future astrophysical simulations. We demonstrate the new algorithm's stability, convergence, and agreement with existing evolution methods. We explain how an innovative spectral approach enables a two orders of magnitude improvement in computational efficiency.
\end{abstract}

\section{What is Characteristic Evolution, and why?}

As an international network of gravitational wave observatories come online, the race to the first direct detection of gravitational waves is expected to herald the beginning of gravitational wave astronomy. Detectors such as Advanced LIGO, VIRGO, GEO, and KAGRA aim for strain sensitivities approaching \(10^{-24}\)\cite{Waldman2011,Accadia:2009zz,Grote:2010zz,Somiya:2012}. Nevertheless, signal candidates from compact binaries or supernovae will be on the cusp of detectability, with very poor signal to noise ratios requiring matched filtering for detection\cite{Allen:2005fk}. Matched filtering requires a comprehensive template bank, the generation of which has been a primary goal of the field of numerical relativity. These templates cover a range of expected astronomical phenomena, and are generated by a variety of numerical codes, including the Spectral Einstein Code (\texttt{SpEC})\cite{Mroue:2013PRL}. Filling out the template bank requires a balance of numerical relativity and analytic waveforms (post-Newtonian\cite{Taracchini:2012} and effective-one-body\cite{Pan:2013rra}), with waveform models often calibrated using numerical results\cite{Bernuzzi:2011aq,Hinderer:2013uwa,Hinder:2013oqa}.

One technical challenge facing the construction of a large template bank is extraction of gauge invariant waveforms from simulations. In general, large computationally intensive simulations are needed to describe the physics of events such as supernovae and compact object binary inspirals. While a waveform-like signal can readily be extracted from anywhere in the computational domain, waveforms are only rigorously defined at future null infinity. Finite radii waveform approximations are universally contaminated by coordinate system dynamics, or gauge effects, which are poorly understood and nearly impossible to remove or even quantify. Comparison with Cauchy Characteristic Extraction, or CCE, an alternative which applies Characteristic Evolution to enable waveform computation at future null infinity, suggests that extrapolation gauge errors could dominate the global error\cite{Taylor:2013zia}. Characteristic Evolution has been previously implemented at up to \(4^{th}\) order radial accuracy\cite{Reisswig:2012ka}, while complete extraction has been achieved with finite difference/volume methods up to \(2^{nd}\) order\cite{Babiuc:2010ze,Oliveira:2011}.

\begin{figure}[h!]
  \centering
    \includegraphics[width=0.7\textwidth]{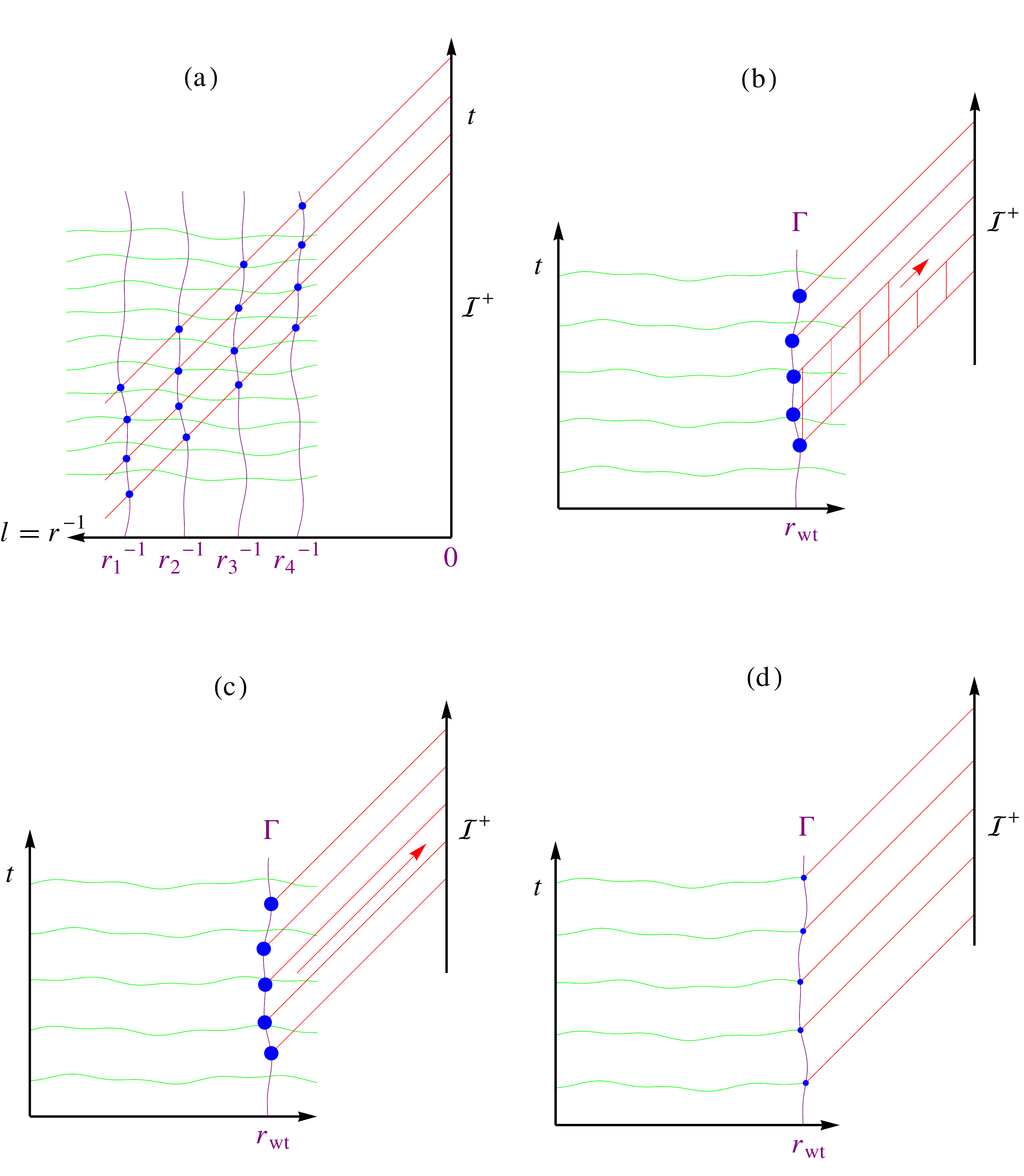}
    \caption{\small{a) Extrapolation methodology. Time-interpolated points (blue) along four nested worldtubes (purple) are derived from metric data on each space-like foliation (green) and used to fit a polynomial in \(l=r^{-1}\), which is extrapolated to \(\mathscr{I}^+\) at \(l=0\). b) The finite difference based characteristic evolution algorithm takes time-interpolated inner boundary conditions and solves each null foliation one null parallelogram at a time in a radial marching algorithm. c) Spectral extraction performs radial integration to \(\mathscr{I}^+\) in a single step. d) A matching algorithm wherein characteristic and Cauchy evolutions share a time parameter and common domain boundary.}}
    \label{fig:ExtrapvsExtract}
\end{figure}

Here we implement inner boundary extraction and evolution. This must be combined with an appropriate outer boundary algorithm to generate the gauge-invariant news at future null infinity. In a hypothetical complete system illustrated in Figure~\ref{fig:ExtrapvsExtract}(d), Cauchy Characteristic Matching (CCM) allows inflowing energy to cross the inner boundary (injection) and enter the Cauchy evolution, ameliorating issues with boundary reflections. Matching, however, faces many technical challenges beyond the scope of this paper, and has only been implemented in the linearized limit\cite{Bishop:1998uk,Szilagyi:2001fy}.

Another method is extrapolation\cite{Taylor:2013zia}, which takes metric quantities at a series of increasing radii, then fits them to a function in \(l = 1/r\). In this coordinate, \(l=0\) corresponds to infinitely far from the origin, where the extrapolated metric quantities are converted to waveforms. Because each sampling point is subject to an unknown degree of coordinate effect contamination, the extrapolated waveform is itself not gauge invariant. 

While characteristic evolution is computationally and conceptually much more involved than extrapolation, it is able to provide gauge invariant waveforms unaffected by coordinate effects. These waveforms are unique modulo the supertranslations, a 4 parameter subgroup of the BMS group, corresponding to arbitrary inertial observer initial conditions at future null infinity\cite{Sachs1962,TamburinoWinicour1966}. Removing gauge effects through characteristic evolution is essential for obtaining accurate and useful waveforms. Essentially, characteristic evolution takes metric data at a topologically spherical worldtube \(\Gamma\) enclosing the relativistic Cauchy evolution and evolves it, as well as initial data, outward to future null infinity, or \(\mathscr{I}^+\). At the outer boundary, the metric quantities can be read off and, in combination with an inertial conformal coordinate system, used to calculate the~true~waveform.

\begin{figure}[h!]
  \centering
    \includegraphics[width=0.8\textwidth]{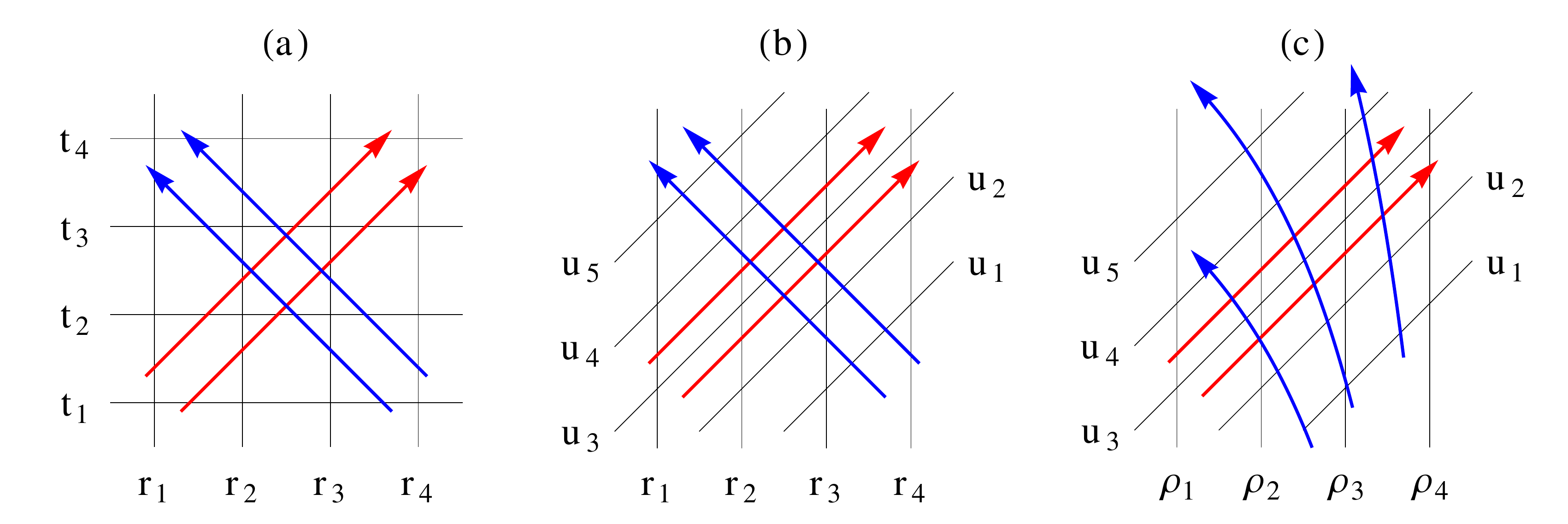}
    \caption{\small{a) Ingoing (blue) and outgoing (red) characteristic worldlines on a radial (\(r\)) grid coordinate system. b) Retarded time (\(u = r+t\)) grid null coordinate parallels outgoing characteristics. c) Compactified null radial (\(\rho = r/(R+r)\)) grid coordinates bring \(\mathscr{I}^+\) into a finite domain. Ingoing characteristics appear curved in this coordinate system.}}
    \label{fig:Transformation}
\end{figure}

The calculation is performed in the Bondi system, in which radial coordinates are outgoing null rays: normal to the worldtube and to each time slice. The spherical coordinates and time-like foliation is adapted from the Cauchy evolution via the worldtube boundary data, illustrated in Figure~\ref{fig:ExtrapvsExtract}(c).

While this simplifies aspects of the evolution, the domain is still infinite, and must be compactified. We use the compactification \(r = R \rho/(1-\rho)\), where \(r\) is the Bondi radius, \(\rho\) is the compactified null radial coordinate and \(R\) the compactification parameter. Setting \(R\) to the Bondi radius of the worldtube \(R=r_{|\Gamma}\) confines \(\rho \in [1/2,1]\), as shown in Figure~\ref{fig:Transformation} (c). While the worldtube is defined by a constant coordinate radius sphere, the Bondi radius at this surface is a variable function of time and angles, giving rise to a wobbly, non-spherical shape in the Bondi coordinate system. This variable compactification parameter requires additional terms in the equations. 

Fixing \(\rho \in [1/2,1]\) and having a variable compactification parameter is different to the approach used in earlier incarnations of CCE, which either interpolated the boundary's variable position or used a different compactification scheme\cite{hpgn,Taylor:2013zia}. While the Pitt null code\cite{Babiuc:2010ze} assumed that a coordinate sphere of constant Cartesian radius formed the worldtube, we lift that constraint here. A fixed computational domain for Characteristic Evolution enables conceptually simple radial integration and dynamically variable extraction radii.

\begin{figure}[h!]
  \centering
    \includegraphics[width=0.9\textwidth]{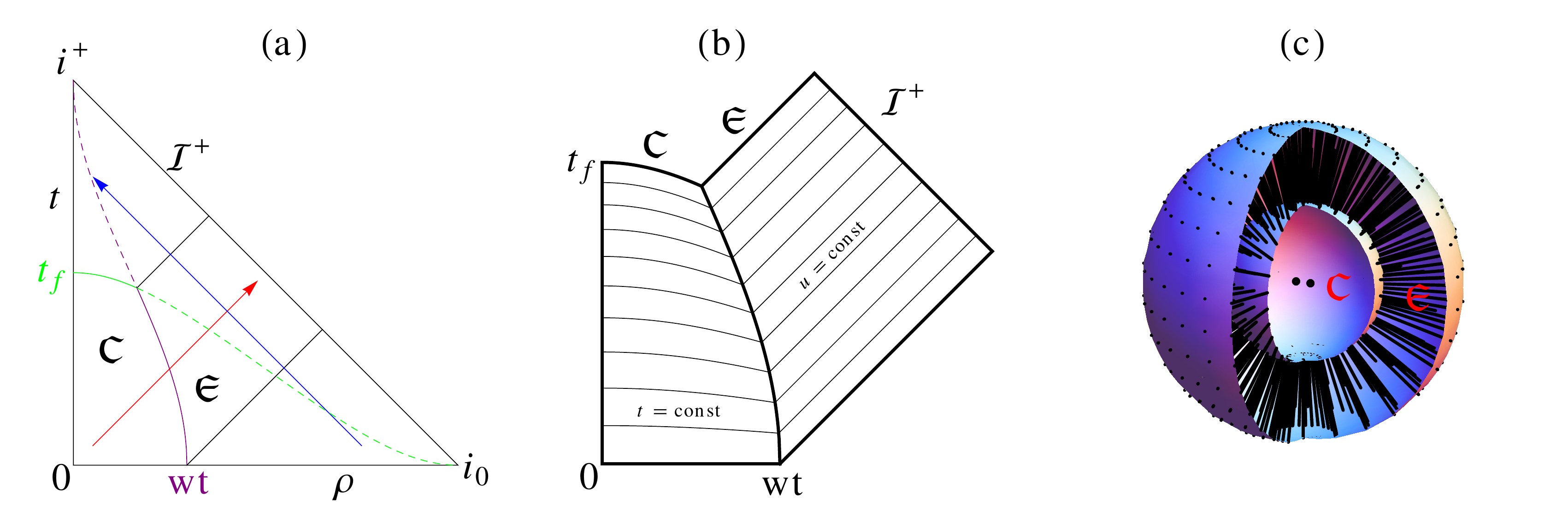}
    \caption{\small{a) Compactified time coordinate yields Penrose-like diagram of global space, depicted here as Minkowski. Space-like foliated Cauchy evolution domain exists in world-sphere \(\mathfrak{C}\), bounded by worldtube `wt'. Null foliated characteristic evolution domain in hollow world-sphere \(\mathfrak{E}\) extends worldtube to \(\mathscr{I}^+\). As before, blue and red arrows represent ingoing and outgoing characteristics respectively. \(t_f\) represents a space-like global final time of Cauchy evolution. Dashed lines extrapolate constant time and radius lines to \(i^+\) and \(i_0\). b) A closer look at the two computational domains with time-like foliation in \(\mathfrak{C}\) and null foliations in \(\mathfrak{E}\) shown for clarity. The worldtube boundary between the two domains is the extraction surface, and doesn't necessary reside on the outer boundary of the Cauchy evolution. c) A 3D rendering showing black radial compactified spokes with the equiangular gridpoint spacing used in our evolution.}}
    \label{fig:Penrose}
\end{figure}

Figure~\ref{fig:Penrose} shows a series of diagrams illustrating the derivation of the computational domain.

\section{Previous Work}
An implementation of Characteristic Evolution was developed by the group of Winicour during the mid 1990s \cite{hpgn,Babiuc:2010ze,Bishop:1998uk}, and is now part of the publicly available Einstein~tool~kit (\texttt{PITTNullCode})\cite{Loffler:2011ay}. In its original form it uses finite differences, achieving \(2^{nd}\) order accuracy. The code has been updated and adapted many times since, and in its current form takes \(\mathcal{O}(\)days\()\) to produce a waveform at an accuracy that matches that of the Cauchy evolution for \(\mathcal{O}(1000M)\) \texttt{SpEC} runs. Five years ago an \(\mathcal{O}(1000M)\) run was sufficiently challenging that there was no CCE bottleneck. Today, with hundreds of runs exceeding dozens of orbits and a few exceeding \(10^5 M\), a faster algorithm is needed. The goal is to have the \texttt{SpEC} characteristic code run alongside the Cauchy evolution at an insigificant additional cost.

This algorithm would build on the formalism of the Pitt null code, but exploit \texttt{SpEC} capabilities to use spectral methods and a much coarser grid to achieve equivalent accuracy. The finite difference algorithm marches outwards one null parallelogram at a time, shown in Figure~\ref{fig:ExtrapvsExtract} (b). A spectral algorithm would rapidly calculate high-accuracy radial data in a single step of numerical integration, as shown in Figure~\ref{fig:ExtrapvsExtract} (c).

The difficulty of a spectral approach lies in consistently treating divergences. As written, the hypersurface evolution equations' source terms are linear or greater order in \(r\) and diverge at infinity. In a finite domain, large but finite terms on either side would numerically cancel, leaving a valid result with no further complications. The Pitt null code uses an asymptotic form to solve the final step to \(\mathscr{I}^+\). To include a point at infinity within a single domain spectral scheme, however, this divergence has to be understood and pre-emptively cancelled. Here, we present a novel approach to regularizing the full nonlinear system, enabling a fully spectral treatment.

\section{A New Algorithm}

\subsection{General Information}

The Bondi metric can be expressed as 
\begin{align}
ds^2 = &-\left(e^{2\beta}(r W+1) - r^2 h_{AB} U^A U^B\right) du^2 \nonumber\\
&- 2 e^{2\beta} du dr - 2 r^2 h_{AB} U^B du dy^A + r^2 h_{AB} dy^A dy^B\;.
\end{align}

\(y^A\) are angular coordinates, where the uppercase Latin indices \(^A\), \(^B\), and so on range from 2 to 3. The quantities \(W, h_{AB}, U^A, Q_A, \beta\) parametrise the metric, representing respectively the mass aspect, the conformal 2-metric, the shift and its radial derivative, and the lapse. Where non-scalar, they are related to complex spin-weighted quantities by contracting them with the appropriate dyad. The dyad \(q^A\) is a complex field on the unit sphere satisfying \(q^Aq_A=0\), \(q^A\bar{q}_A = 2\), \(q^A = q^{AB}q_B\), with \(q^{AB}q_{BC} = \delta^A_C\) and \(q_{AB} = (q_A\bar{q}_B + \bar{q}_A q_B)/2\), the unit sphere metric. Under this convention, the spin-weighted functions \(U = U^A q_A\) and \(Q = Q_A q^A\), while \(J = h_{AB}q^A q^B/2\) uniquely determines the spherical conformal 2-metric component of the general 4-metric\cite{hpgn}. We chose a dyad consistent with our formulation of the eth operator \(\eth\)\cite{Gomez:1997}, given by \(q^A=(-1,-i/\sin \theta)\), which is regular everywhere except the poles, which we can avoid through careful choice of grid points. It is worth noting that any choice of angular coordinates are possible. Other conventions use multiple patches to avoid singularities at the poles, in which the phase dislocation due to spin-weight when moving from patch to patch is explicit.

The key to the Bondi formulation is that all the spin-weighted metric quantities have a heirarchical structure, enabling their natural ordering as a nested series of self-referential equations on the outgoing null hypersurface. Their derivation is discussed in Bishop {\it et al.}\cite{Bishop:1998uk,hpgn}. In a relatively simple form they are

\begin{align}
\beta,_r = & N_\beta\;,\\
(r^2 Q),_r = & -r^2(\bar{\eth} J + \eth K),_r + 2r^4 \eth(r^{-2}\beta),_r + N_Q\;,\\
U,_r = & r^{-2} e^{2\beta} Q + N_U\;,\\
(r^2W),_r = & \frac{1}{2} e^{2\beta} \mathcal{R} -1 - e^\beta \eth \bar{\eth} e^\beta + \frac{1}{4} r^{-2} (r^4(\eth \bar{U} + \bar{\eth} U )),_r + N_W\;,\\
&\mathrm{where}\;\mathrm{the}\;\mathrm{scalar}\;\mathrm{curvature}\\
\mathcal{R} = & 2K - \eth \bar{\eth} K + \frac{1}{2}(\bar{\eth}^2 J + \eth^2 \bar{J}) + \frac{1}{4K} (\bar{\eth} \bar{J} \eth J - \bar{\eth}J \eth \bar{J})\;,\\
&\mathrm{and}\;\mathrm{the}\;\mathrm{time}\;\mathrm{derivative}\;\mathrm{term}\\
2(r J),_{ur} = & \left((1+r W )(rJ),_r\right),_r -r^{-1}(r^2 \eth U),_r + 2 r^{-1}e^\beta \eth^2 e^\beta - (r W),_r J + N_J\;,\\
&\mathrm{where}\\
1 = & K^2 - J \bar{J}. 
\end{align}

The nonlinear terms \(N_\beta, N_Q, N_U, N_W, N_J\) are given in Appendix~D\footnote{Note that \(W\) is defined according to the convention in \cite{Bishop:1998uk}, which differs from \cite{hpgn} by a factor of \(r^2\).}. The radial compactification is given by \(r = R \rho/(1-\rho)\), where the compactification parameter \(R(u,\theta,\phi)\) is the (not necessarily constant) Bondi areal radius of the worldtube.

On each hypersurface in the spacetime foliation, each equation is solved in turn. Given \(J\), \(\beta\) is solved, then \(U\), \(Q\), and \(W\) in turn, enabling the computation of \(J,_u\). \(J,_u\) permits a step forward in time and \(J\) is thus defined on the next hypersurface.

Spherical derivatives are implemented using the eth operator \(\eth\) on spin-weighted spherical harmonics. \(\eth\) is given by the contraction of the dyad with the derivative operator\cite{Gomez:1997}.

In spherical coordinates, this can be written 
\begin{align}
\eth\eta = -(\sin^s \theta) \left(\frac{\partial}{\partial \theta} + \frac{i}{\sin\theta}\frac{\partial}{\partial \phi}\right)(\sin^{-s}\theta\, \eta)\;,\\
\bar{\eth}\eta = -(\sin^{-s} \theta) \left(\frac{\partial}{\partial \theta} - \frac{i}{\sin\theta}\frac{\partial}{\partial \phi}\right)(\sin^s\theta\, \eta)\;.
\end{align}
\(\eth\) and \(\bar{\eth}\) increment and decrement respectively the spin-weight \(s\) of the quantity they act upon. Details are given in Appendix~C.

In the Pitt null code approach to CCE, a finite difference-based algorithm is used to solve the hypersurface equations, radially marching from the inner boundary outward, one point at a time. In our algorithm, we use spectral methods to calculate the radial indefinite integral on all collocation points along a spherical radial spoke in a single calculation. A spectral approach is faster and more accurate, but requires the finessing of a few technical difficulties.

The most obvious of these appears in the hypersurface equations for~\(W\) and~\(Q\). To calculate a numerical integral, it is necessary to express the integrand as a bounded function on all radial collocation points, including those at \(\mathscr{I}^+\). Given that \(Q\) is regular and \(\mathcal{O}(1)\) at \(\mathscr{I}^+\), \(r^2 Q\) is clearly divergent. The integral of the right hand side is similarly divergent for the outermost collocation point, where \(r \rightarrow \infty\) or \(\rho \rightarrow 1\). Here, we solve this problem by expressing the right hand side as a Laurent series around the pole at \(\rho = 1\) and then repeatedly integrating by parts. Mathematically, if the equation is written 
\begin{equation} 
(r^2 Q)_{,\rho} = A + \frac{B}{1-\rho} + \frac{C}{(1-\rho)^2} + \frac{D}{(1-\rho)^3}\;, 
\end{equation}
then, as shown in Appendix A,
\begin{align}
Q = &Q_0 + \frac{D}{2 \rho^2 R^2} - (-C+D'/2)\frac{1-\rho}{\rho^2 R^2} - (B - C' + D''/2) \frac{(1-\rho)^2 \mathrm{log}(1-\rho)}{\rho^2 R^2} \nonumber\\
- &(B' - C'' + D'''/2)\frac{(\rho+(1-\rho)\mathrm{log}(1-\rho))(1-\rho)^2}{\rho^2 R^2} \nonumber\\
+ &\frac{(1-\rho)^2}{\rho^2 R^2} \int \; A + (B'' -C''' + D''''/2)(\rho+(1-\rho)\mathrm{log}(1-\rho))\;, 
\end{align}
where \(D' = \frac{\partial D}{\partial \rho}\) and so on. In this context, the \(\int\) operator refers to a radial numerical integral where the integration constant is fixed such that the inner boundary value (at the worldtube) vanishes. Note that all terms within the integral are bounded within the domain \(\rho \in [1/2,1]\), including at \(\mathscr{I}^+\). On the outer boundary, terms in \(\log(1-\rho)\) must cancel out to preserve \(C^\infty\) differentiability.

A second, less obvious, issue is that the right hand side of the \(J,_{u \rho}\) hypersurface equation has nonlinear terms with the desired quantity \(J,_u\) in them (as seen in the equation for \(P_1\), Eqn A6, Bishop {\it et al.} \cite{Bishop:1998uk}). In order to perform the radial integral and solve for \(J,_u\) in a single step, these nonlinear terms have to be somehow removed. One approach is to factorise using an integrating factor and, conceptually, that is what is done. We shall illustrate this first with a simple example.

Given 

\begin{align}
J,_{u\rho} + A J,_u = B\;, \; \mathrm{then} \\
\left(e^{\int A} J,_u\right),_\rho = e^{\int A} B\;,\;\mathrm{and}\\
J_{,u} =  e^{-\int A}\int e^{\int A} B\;.
\end{align}

The actual equations are, however, more difficult as the variable is complex, and the nonlinear term also includes the complex conjugate \(\bar{J},_u\) term, which makes a simple factorization impossible. Writing \(J,_u = \Phi\), the actual equation can be written

\begin{equation}
(r \Phi)_{,\rho} - (r J) (\Phi \bar{\Gamma} + \bar{\Phi} \Gamma) = A + \frac{B}{1-\rho} + \frac{C}{(1-\rho)^2}\;, \; \mathrm{where}
\end{equation}
\begin{equation}
\Gamma = \left(J_{,\rho} - J \frac{K_{,\rho}}{K}\right)\;.
\end{equation}

The right hand side is a Laurent expansion analogous to the equations for Q and W. The part \((\Phi \bar{\Gamma} + \bar{\Phi} \Gamma)\) is a quantity plus its conjugate and is thus wholey real. This leads to the insight that a real-imaginary split of the equation is productive. Writing \(J = J_\mathcal{R} + i J_\mathcal{I}\), we can exploit the isomorphism between complex numbers and non-singular \(2\times 2\) matrices by writing

\begin{equation}
\begin{pmatrix} r \Phi_\mathcal{R} \\ r \Phi_\mathcal{I} \end{pmatrix},\rho
-
\begin{pmatrix} J_\mathcal{R} \Gamma_\mathcal{R} & J_\mathcal{R} \Gamma_\mathcal{I} \\
  J_\mathcal{I} \Gamma_\mathcal{R} & J_\mathcal{I} \Gamma_\mathcal{I}
\end{pmatrix}
\begin{pmatrix} r \Phi_\mathcal{R} \\ r \Phi_\mathcal{I} \end{pmatrix}
=
\begin{pmatrix} RHS_\mathcal{R} \\ RHS_\mathcal{I} \end{pmatrix}\;.
\end{equation}

We have restored the integrating factor form of the equation, only now in matrix form. The use of non-constant matrices for an integrating factor requires the calculation of commutators, which in all but the trivial case are non-zero. The required formalism is the Magnus expansion\cite{Magnus1954,MagnusReview2009}, in which the usual integrating factor is supplemented by integrals of progressively higher order commutators.

Let
\begin{equation}
F
=
\begin{pmatrix} J_\mathcal{R} \Gamma_\mathcal{R} & J_\mathcal{R} \Gamma_\mathcal{I} \\  J_\mathcal{I} \Gamma_\mathcal{R} & J_\mathcal{I} \Gamma_\mathcal{I}
\end{pmatrix}\;,
\end{equation}
and
\begin{equation}
{\bf \Phi}
=
\begin{pmatrix} \Phi_\mathcal{R} \\  \Phi_\mathcal{I}
\end{pmatrix}\;.
\end{equation}
Then 
\begin{equation}
(r {\bf \Phi})_{,\rho} - F . (r {\bf \Phi}) = \mathrm{exp} \big(\Omega(\rho)\big).\Big(\mathrm{exp} \big(-\Omega(\rho)\big) . (r {\bf \Phi})\Big)_{,\rho} \;,
\end{equation}
where
\begin{equation}
\Omega(\rho) = \sum_{k=1}^\infty \Omega_k(\rho)\;.
\end{equation}

Write \(F(\rho_1) = F_1\). Then \(\Omega_k\) forms a series called the Magnus expansion.
\begin{align}
\Omega_1(\rho) &= \int_{\rho_0}^{\rho} d\rho_1 F_1\;,\\
\Omega_2(\rho) &= \frac{1}{2} \int_{\rho_0}^{\rho} d\rho_1 \int_{\rho_0}^{\rho} d\rho_2 [F_1,F_2]\;,\\
\Omega_3(\rho) &= \frac{1}{6} \int_{\rho_0}^{\rho} d\rho_1 \int_{\rho_0}^{\rho} d\rho_2 \int_{\rho_0}^{\rho} d\rho_3 [F_1,[F_2,F_3]]+[F_3[F_2,F_1]]\;,
\end{align}
and so on.

In practise, this series must be truncated. Fortunately, \(|F|\approx 10^{-3}<<1\) in practical cases, so the series converges rapidly, necessitating calculation of only the first three terms. Although a formalism exists\cite{TamburinoWinicour1966} to deal with the non-linear terms without resorting to the Magnus expansion, it requires transformation in terms of dyads, whose issues around the poles we have been careful to avoid. 

With the Magnus expansion, the troublesome nonlinear terms are readily dealt with and the equation can be expressed in the form

\begin{equation} 
(r \Phi),_\rho = \tilde{A} + \frac{\tilde{B}}{1-\rho} + \frac{\tilde{C}}{(1-\rho)^2}\;.
\end{equation}

This is solved analogously to the radial hypersurface equations for \(W\) and~\(Q\).

\subsection{Details specific to our implementation}

In our implementation, we used a spherical coordinate system. The Chebyshev pseudospectral method was used for the (1D interval) radial basis function. Spherepack (for real tensor quantities) and Spinfast (for complex spin-weighted quantities) were used for the 2D spherical basis function\cite{Boyd1989,spherepack-home-page,Huffenberger:2010}. 

Calculus operations such as integration, differentiation, and computation of \(\eth\) were performed in each basis function according to standard methods. Time stepping was performed with an adaptive Dormand Prince \(5^{th}\) order routine, or Runge-Kutta 4 with constant time steps. Spectral filtering of spatial quantities ensured that all system modes remained within the time integrators' domains of stability. Specifically, the \(i^{th}\) radial coefficient was filtered by a factor \(exp(-108 (i/(N_\rho-1))^{24})\), where \(i\) varies between 0 and \(N_\rho-1\). Angular coefficients were set to zero for \(l=L_{max}\) and \(l=L_{max}-1\). Filtering was applied at the end of each time step and to the hypersurface quantites \(Q\), \(W\), and \(H\) after their computation.

\section{Stability and Convergence}

The spectral characteristic evolution algorithm is analytically derived, but how does it actually perform? We tested the stability and convergence of the code under a variety of circumstances designed to far exceed the demands of any actual CCE run\cite{Taylor:2013zia}.

\begin{figure}[h!]
  \centering
    \includegraphics[width=0.8\textwidth]{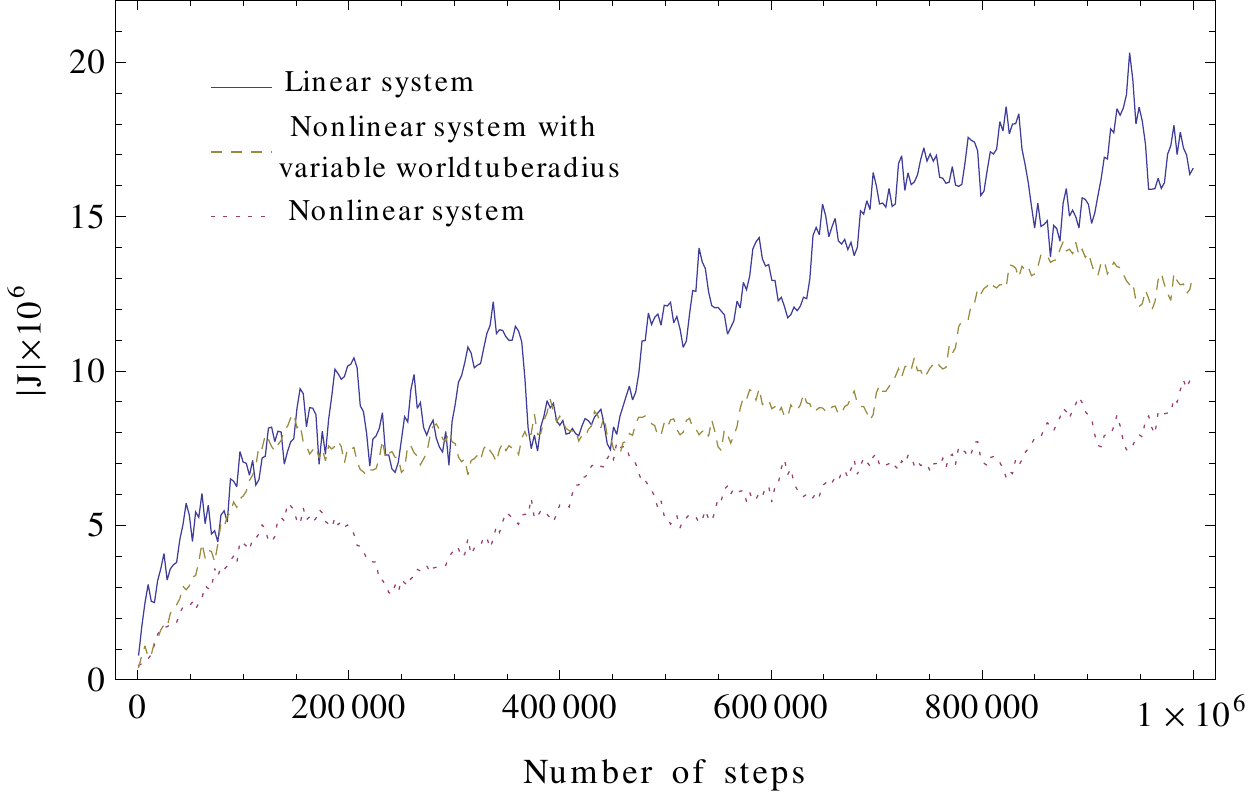}
    \caption{\small{Graph of \(|J|\) over a million time steps with a white noise boundary conditions. The three lines represent a linear baseline, a nonlinear run, and the full nonlinear system including boundary position variation. All three runs display sub-exponential growth, indicating stability. For the linear test, a minimal resolution of \(N_\rho=4\), \(L = 8\) was used. For the nonlinear tests, a minimal resolution of \(N_\rho=8\), \(L = 12\) was used. All tests used an adaptive RK3 time stepper.}}
    \label{fig:StabTest}
\end{figure}

For a stability test, we ran the algorithm with a variety of settings for a million steps with white noise initial and boundary conditions of magnitude \(10^{-6}\). The linear setting truncates all nonlinear terms, and represents a baseline condition. The nonlinear setting restores all nonlinear terms in the equations. The most general setting includes a variable inner boundary position, encoded in the magnitude of the compactification parameter \(R\). Each of these three conditions was run, and in all three cases, the norm of \(J\) was stable. In particular, all three runs do not grow exponentially, as seen in Figure~\ref{fig:StabTest}.

\begin{figure}[h!]
  \centering
    \includegraphics[width=0.8\textwidth]{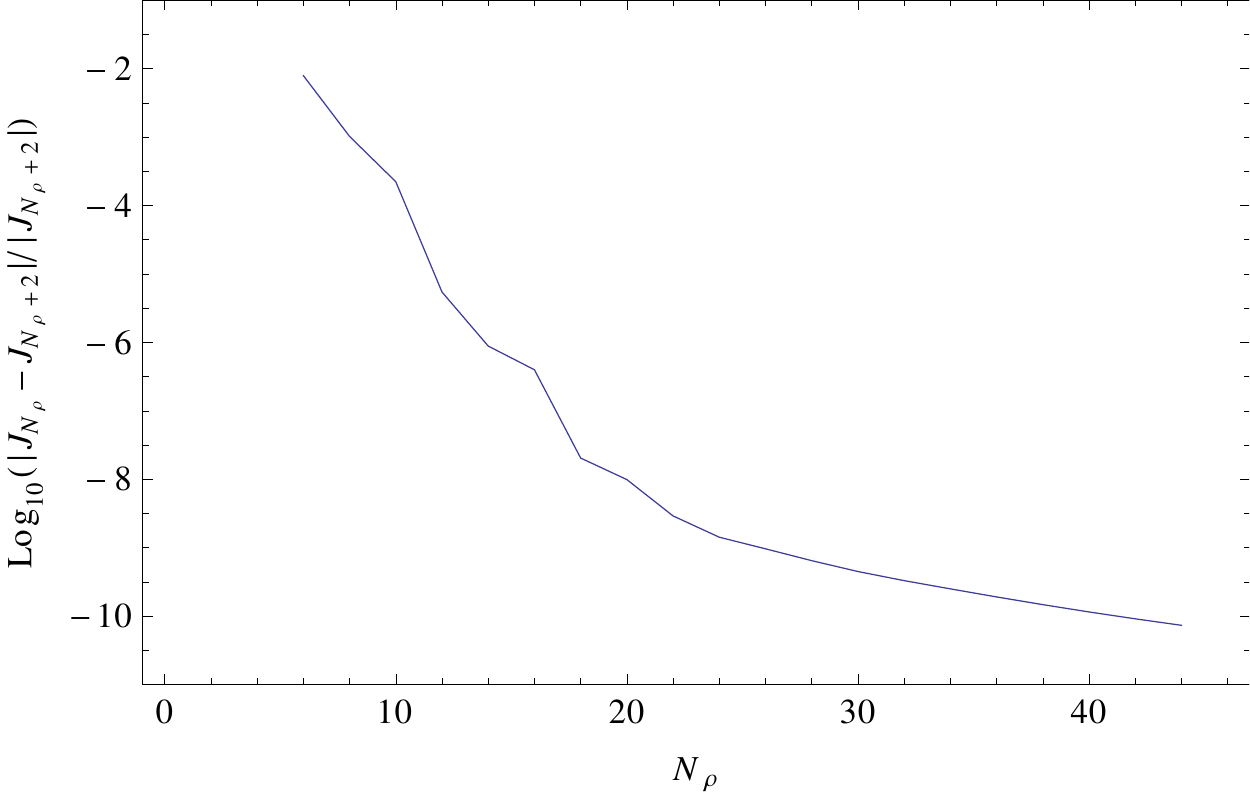}
    \caption{\small{Log(relative error) shows a linear relationship with number of radial points \(N_\rho\), indicating spectral convergence. At \(N_\rho \approx 24\), convergence turns sub-exponential. Error is calculated according to \(\underset{\small{1050M<t<1100M}}{\operatorname{Mean}}\left(\log_{10}\frac{|J_{N_\rho}-J_{N_\rho+2}|_\infty}{|J_{N_\rho+2}|_\infty}\right)\) for the generic precessing run. For these runs, \(L = 16\), \(\Delta t = 0.4M\), RK4 is used, and \(N_\rho\) varies from 6 to 46.}}
    \label{fig:RadConv}
\end{figure}

The purpose of a spectral algorithm is to obtain faster convergence, particularly with respect to radial integration. We ran a series of tests while varying \(N_\rho\) between 6 and 46. The test was run on the generic run discussed in Section 6 between \(t=1000M\) and \(t=1100M\), and the results averaged between \(t=1050M\) and \(t=1100M\) to remove any transient contamination. While our algorithm can use any one-dimensional spectral representation for the radial direction, these tests were conducted using Gauss Chebyshev Lobatto polynomial basis functions, as discussed in the previous section. Figure~\ref{fig:RadConv} shows that local relative error in \(J\) converges exponentially as the number of radial points increases up to around 24, at which point convergence becomes sub-exponential due to roundoff error introduced in the integration algorithm.

\begin{figure}[h!]
  \centering
    \includegraphics[width=0.8\textwidth]{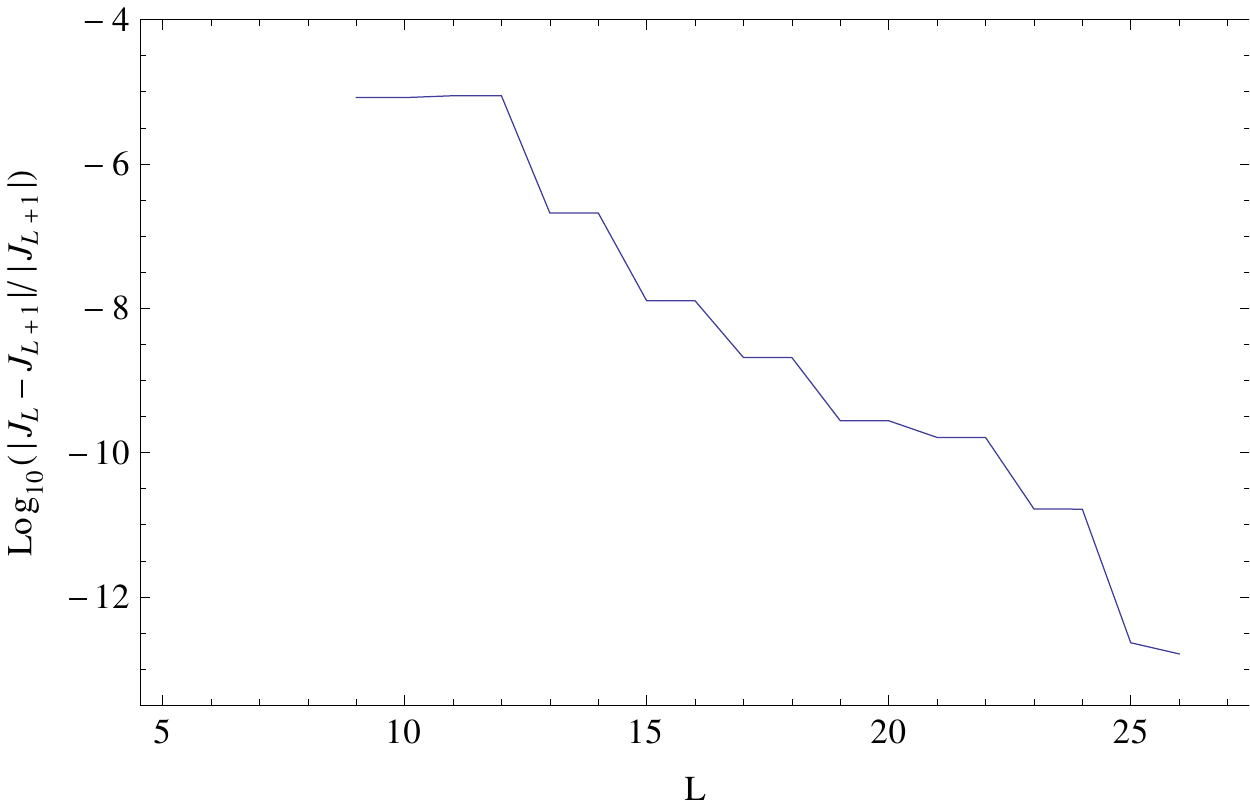}
    \caption{\small{Log(relative error) vs spherical resolution shows rapid spectral convergence. Error is calculated analogously to Figure~\ref{fig:RadConv}. For these runs, \(N_\rho = 24\), \(\Delta t = 0.4M\), RK4 is used, and \(L\) varies from 8 to 27.}}
    \label{fig:AngConv}
\end{figure}

Convergence with angular resolution was calculated identically to that for Figure~\ref{fig:RadConv}, and shows rapid exponential convergence, as seen in Figure~\ref{fig:AngConv}. Our implementation uses the Spinsfast package\cite{Huffenberger:2010} for spin-weighted spherical harmonic computation in a manner analogous to Spherepack.

\begin{figure}[h!]
  \centering
    \includegraphics[width=0.8\textwidth]{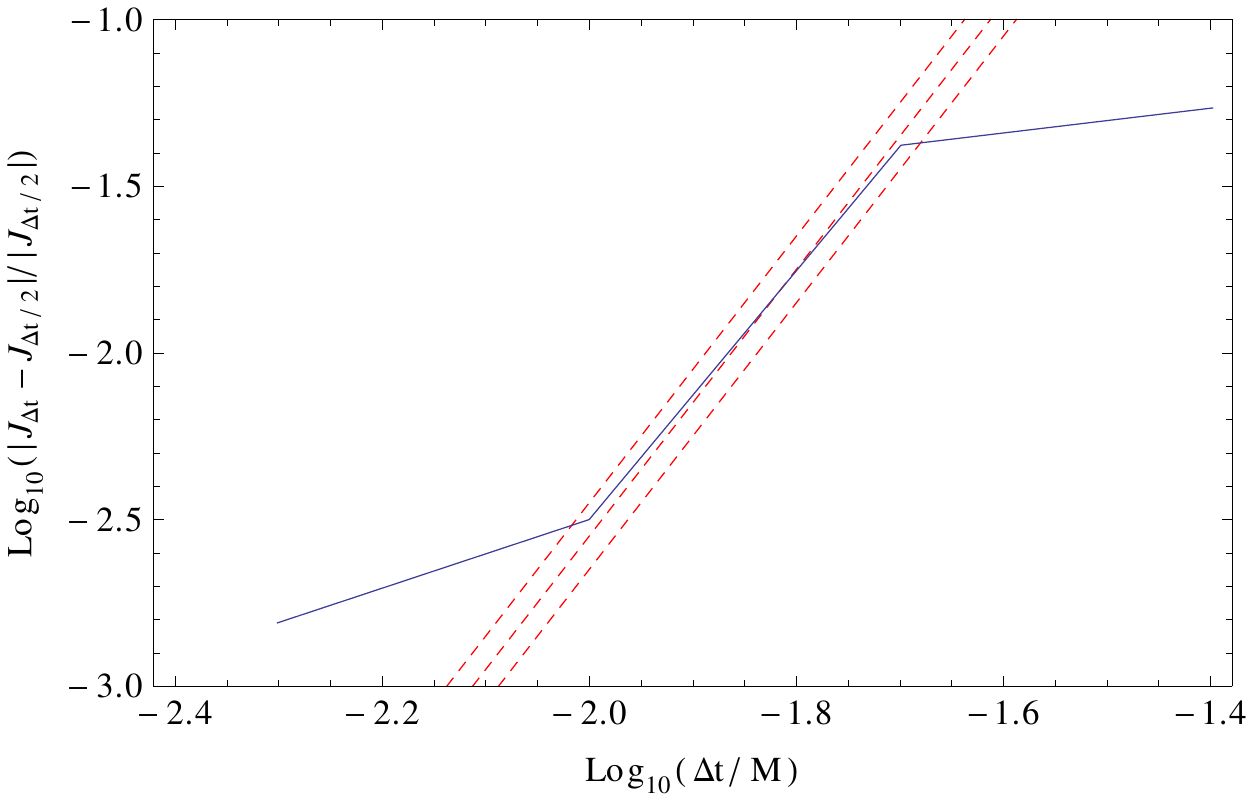}
    \caption{\small{Log plot of relative error vs time step shows \(4^{th}\) order convergence, demarcated by parallel dashed red lines. Error is calculated analogously to Figure~\ref{fig:RadConv}. For this test the minimum grid spacing is adjusted with time step to maintain a constant ratio. For time steps of \((0.04M,0.02M,0.01M,0.005M,0.0025M)\), the number of radial points was \((6, 8, 11, 15, 21)\) respectively. Angular resolution was held constant with \(L = 9\).}}
    \label{fig:TimeStepConv}
\end{figure}

Finally, timestep convergence was analysed. For the purposes of this test, the minimum grid spacing to timestep was held constant, while the timestep was varied over more than an order of magnitude. The code displays \(4^{th}\) order time convergence, consistent with the chosen integrator RK4, as shown in Figure~\ref{fig:TimeStepConv}.

\section{Comparison With Finite Differences Evolution}
\begin{figure}[h!]
  \centering
    \includegraphics[width=0.9\textwidth]{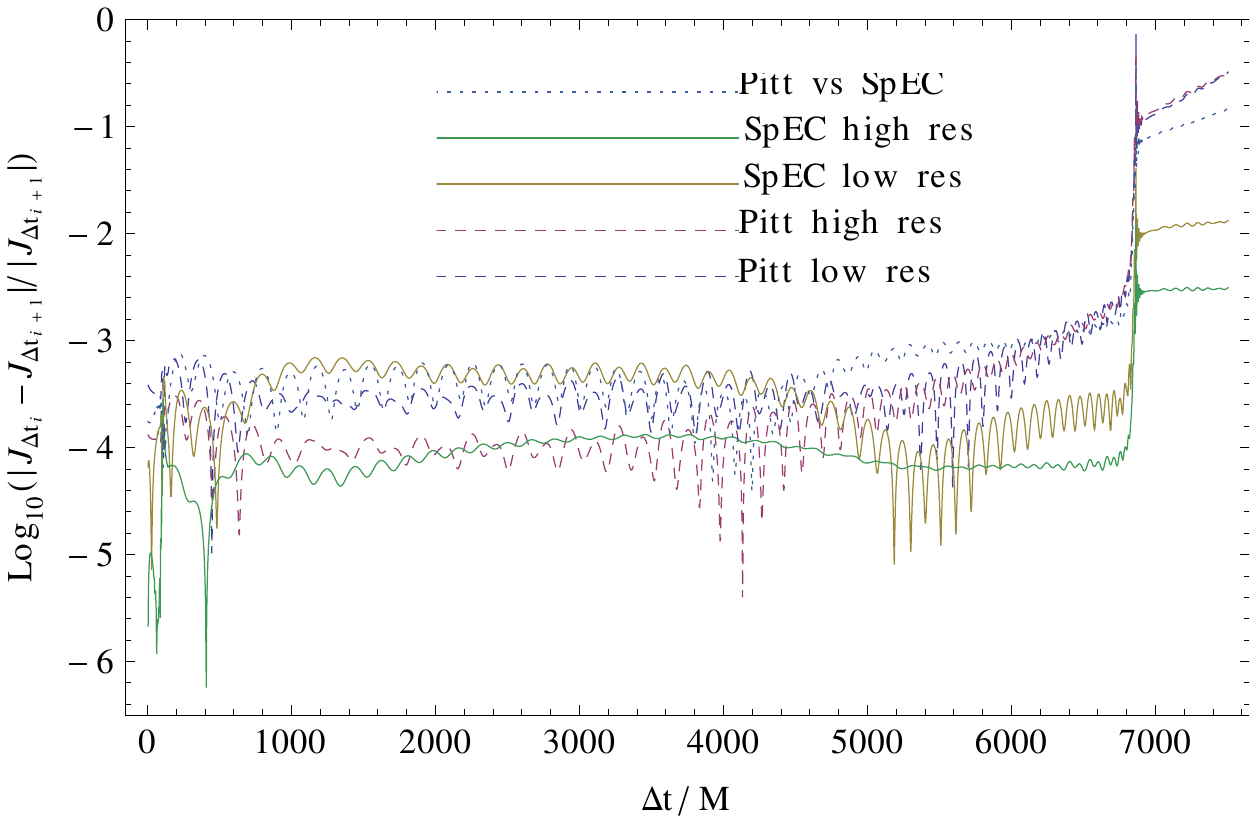}
    \caption{\small{Comparison of convergence of Pitt and \texttt{SpEC} evolution codes. The Pitt runs (dashed) show \(2^{nd}\) order convergence until around \(5000M\), when convergence is gradually lost. The \texttt{SpEC} runs (solid), with parameters chosen for comparible levels of error, display higher order time convergence throughout the entire run. The dotted line shows the level of agreement between the highest resolution runs of each code, consistent with their respective resolution of junk radiation. Peak amplitude \(J,_{uu}\) occurs at \(t = 6832M\), at which point error in relative amplitude and phase is \(10^{-2.367}\) and \(0.002\) respectively. These values represent expected error for a strain calculation.}}
    \label{fig:FullRunConv}
\end{figure}

For longer run comparisons we used the generic precessing Binary Black Hole simulation detailed in Table 1 (case 4) of Taylor {\it et al.}\cite{Taylor:2013zia}. Its parameters are mass ratio \(q=3\), black hole spin \(\chi_1 = (0.7,0,0.7)/\sqrt{2}\) and \(\chi_2 = (-0.3,0,0.3)/\sqrt{2}\), number of orbits 26, total time \(T=7509M\), initial eccentricity \(10^{-3}\), initial frequency \(\omega_{ini} = 0.032/M\), and extraction (coordinate) radius \(R=100M\). We performed 3 runs using the PITT null code for baseline comparisons with 3 runs using the new spectral Characteristic Evolution code, using parameters shown in Table~\ref{tab:Params}. The spectral code converges rapidly to within the error implied by the PITT null code, as shown in Figure~\ref{fig:FullRunConv}. Parameters for the \texttt{SpEC} runs were specifically chosen for comparible levels of error with the Pitt null code. Note that despite \texttt{SpEC} code spatial resolution parameters being chosen for consistent minimum grid spacing to time step ratio, rather than equal numbers of points, the global error comparison is not adversely affected. Both codes were run on the same cluster with dense output.

\begin{table}
\centering
  \begin{tabular}{| r | c c c | c c c |}
    \hline
    Run & Pitt 1 & Pitt 2 & Pitt 3 & \texttt{SpEC} 1 & \texttt{SpEC} 2 & \texttt{SpEC} 3 \\
    \hline
    \(N_r\) & 100 & 150 & 200 & 10 & 12 & 14 \\
    \(N_{stereo}\) or \(L\) & 40 & 60 & 80 & 12 & 14 & 17 \\
    \(\Delta t/M\) & 0.1 & \(0.0666\dots\) & 0.05 & 1.0 & \(0.666\dots\) & 0.5 \\
    T (CPU hours) & 2688 & 5760 & 6912 & 12 & 31 & 52 \\
    \hline
  \end{tabular}
  \caption{\small{Parameters used for code comparisons. The Pitt null code uses two stereographic patches of size \(N_{stereo}^2\). For both codes, the total number of angular points is given by \(2 N_{stereo}^2\) and \(2 L^2\) respectively.}}
  \label{tab:Params}
\end{table}

Note also that the resolution of the \texttt{SpEC} runs is an order of magnitude lower, the time steps an order of magnitude longer, and the run time more than two orders of magnitude faster for equivalent or superior accuracy. In this case, accuracy is ultimately limited by the order of the time stepper (RK4) and the order of the time interpolation on the inner boundary (also \(4^{th}\) order).

Global accuracy and convergence was assessed by graphing the relative complex difference between runs of adjacent accuracy. The errors are computed according to 
\[ E_{Pitt\; low\; res} = \frac{|j_{{22}_{Pitt\; 1}}-j_{{22}_{Pitt\; 2}}|}{|j_{{22}_{Pitt\; 2}}|},\]
\[ E_{Pitt\; high\; res} = \frac{|j_{{22}_{Pitt\; 2}}-j_{{22}_{Pitt\; 3}}|}{|j_{{22}_{Pitt\; 3}}|},\]
\[ E_{SpEC\; low\; res} = \frac{|j_{{22}_{SpEC\; 1}}-j_{{22}_{SpEC\; 2}}|}{|j_{{22}_{SpEC\; 2}}|},\]
\[ E_{SpEC\; high\; res} = \frac{|j_{{22}_{SpEC\; 2}}-j_{{22}_{SpEC\; 3}}|}{|j_{{22}_{SpEC\; 3}}|},\]
\[ E_{Pitt\; vs\; SpEC} = \frac{|j_{{22}_{SpEC\; 3}}-j_{{22}_{Pitt\; 3}}|}{|j_{{22}_{SpEC\; 3}}|},\]
where \(j_{22}\) is the \(_{22}\) spherical harmonic coefficient of \(J\).

\begin{figure}[h!]
  \centering
    \includegraphics[width=0.9\textwidth]{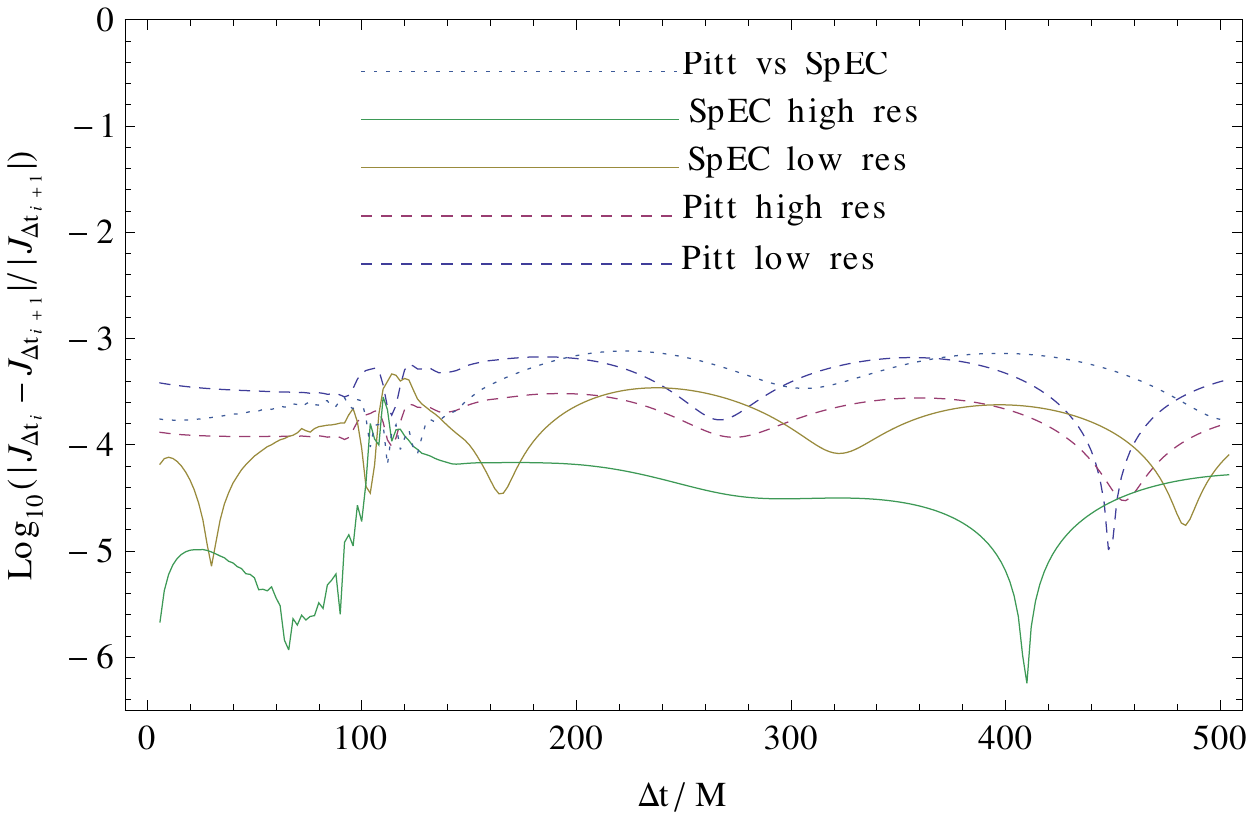}
    \caption{\small{First \(500M\) of Figure~\ref{fig:FullRunConv} shows very high initial agreement of codes until evolution of junk radiation. Nevertheless, agreement remains consistently high.}}
    \label{fig:FullRunConvStart}
\end{figure}

Figure~\ref{fig:FullRunConv} shows strong and consistent convergence of the \texttt{SpEC} runs over the full \(7509M\), as well as consistent agreement between the \texttt{SpEC} and Pitt runs. Figure~\ref{fig:FullRunConvStart} shows convergence during the junk radiation part of the run, where partial loss of agreement between the Pitt and \texttt{SpEC} runs is caused by their respective differences in unphysical junk radiation resolution. The error nevertheless remains well bounded for the entire run.

\section{Conclusion}

A new algorithm for spectral Characteristic Evolution has been developed, implemented, and demonstrated within the \texttt{SpEC} framework. It exploits analytic regularization of the vacuum hypersurface equations and the accuracy and speed of spectral methods. Stability, self-convergence, and convergent agreement with the existing finite difference Characteristic Evolution code are demonstrated. This algorithm will form the basis for a complete extraction and matching methodology that will enable gauge invariant waveforms and junk radiation-free initial conditions to be computed on the fly.

\ack{
  We thank Jeffrey Winicour for his invaluable insight and encyclop\ae dic knowledge of all things Extraction. We thank Nicholas Taylor for his generic spin BBH run we used to test and baseline code performance. We thank Mark Scheel, Yanbei Chen, and Christian Reisswig for their advice, support, and technical expertise. This research used the Spectral Einstein Code (\texttt{SpEC})\cite{Mroue:2013PRL}. The Caltech cluster \texttt{zwicky.cacr.caltech.edu} is an essential resource for \texttt{\texttt{SpEC}} related research, supported by the Sherman Fairchild Foundation and by NSF award PHY-0960291. This research also used the Extreme Science and Engineering Discovery Environment (XSEDE) under grant TG-PHY990002. The UCSD cluster \texttt{ccom-boom.ucsd.edu} was used during code development. This project was supported by the Sherman Fairchild Foundation, and by NSF Grants PHY-1068881, AST-1333520, and CAREER Grant PHY-0956189.

\section*{References}

\bibliographystyle{utphys}
\bibliography{References/References}

\providecommand{\href}[2]{#2}\begingroup\raggedright\begin{thebibliography}{10}

\bibitem{Waldman2011}
S.~J. Waldman, ``The advanced ligo gravitational wave detector,'' Tech. Rep.
  LIGO-P0900115-v2, {LIGO} Project, 2011.

\bibitem{Accadia:2009zz}
T.~Accadia, F.~Acernese, F.~Antonucci, P.~Astone, G.~Ballardin, {\em et~al.},
  \href{http://dx.doi.org/10.1142/9789814374552_0313}{``{Plans for the upgrade
  of the gravitational wave detector VIRGO: Advanced VIRGO},''} in {\em
  Proceedings of the Twelfth Marcel Grossmann Meeting on General Relativity},
  T.~Damour, R.~T. Jantzen, and R.~Ruffini, eds., pp.~1738--1742.
\newblock 2009.

\bibitem{Grote:2010zz}
{\bfseries LIGO Scientific Collaboration} Collaboration, H.~Grote, ``{The GEO
  600 status},''
\href{http://dx.doi.org/10.1088/0264-9381/27/8/084003}{{\em Class.\ Quantum
  Grav.} {\bfseries 27} (2010) 084003}.

\bibitem{Somiya:2012}
K.~Somiya and the {KAGRA}~Collaboration, ``Detector configuration of
  {KAGRA}--the japanese cryogenic gravitational-wave detector,''
  \href{http://dx.doi.org/10.1088/0264-9381/29/12/124007}{{\em Class.\ Quantum
  Grav.} {\bfseries 29} no.~12, (2012) 124007}.

\bibitem{Allen:2005fk}
B.~Allen, W.~G. Anderson, P.~R. Brady, D.~A. Brown, and J.~D. Creighton,
  ``{FINDCHIRP: An Algorithm for detection of gravitational waves from
  inspiraling compact binaries},''
  \href{http://dx.doi.org/10.1103/PhysRevD.85.122006}{{\em Phys.Rev.}
  {\bfseries D85} (2012) 122006},
\href{http://arxiv.org/abs/gr-qc/0509116}{{\ttfamily arXiv:gr-qc/0509116
  [gr-qc]}}.

\bibitem{Mroue:2013PRL}
A.~H. Mroue, M.~A. Scheel, B.~Szilagyi, H.~P. Pfeiffer, M.~Boyle, D.~A.
  Hemberger, L.~E. Kidder, G.~Lovelace, S.~Ossokine, N.~W. Taylor,
  A.~Zenginoglu, L.~T. Buchman, T.~Chu, E.~Foley, M.~Giesler, R.~Owen, and
  S.~A. Teukolsky, ``A catalog of 174 binary black hole simulations for
  gravitational wave astronomy,''
  \href{http://dx.doi.org/10.1103/PhysRevLett.111.241104}{{\em Phys.\ Rev.\
  Lett.} {\bfseries 111} (2013) 241104},
  \href{http://arxiv.org/abs/1304.6077}{{\ttfamily arXiv:1304.6077 [gr-qc]}}.

\bibitem{Taracchini:2012}
A.~{Taracchini}, Y.~{Pan}, A.~{Buonanno}, E.~{Barausse}, M.~{Boyle}, T.~{Chu},
  G.~{Lovelace}, H.~P. {Pfeiffer}, and M.~A. {Scheel}, ``{Prototype
  effective-one-body model for nonprecessing spinning inspiral-merger-ringdown
  waveforms},'' \href{http://dx.doi.org/10.1103/PhysRevD.86.024011}{{\em Phys.\
  Rev.\ D} {\bfseries 86} no.~2, (July, 2012) 024011},
  \href{http://arxiv.org/abs/1202.0790}{{\ttfamily arXiv:1202.0790 [gr-qc]}}.

\bibitem{Pan:2013rra}
Y.~Pan, A.~Buonanno, A.~Taracchini, L.~E. Kidder, A.~H. Mroue, {\em et~al.},
  ``{Inspiral-merger-ringdown waveforms of spinning, precessing black-hole
  binaries in the effective-one-body formalism},''
  \href{http://arxiv.org/abs/1307.6232}{{\ttfamily arXiv:1307.6232 [gr-qc]}}.
arXiv:1307.6232.

\bibitem{Bernuzzi:2011aq}
S.~Bernuzzi, M.~Thierfelder, and B.~Bruegmann, ``{Accuracy of numerical
  relativity waveforms from binary neutron star mergers and their comparison
  with post-Newtonian waveforms},''
  \href{http://dx.doi.org/10.1103/PhysRevD.85.104030}{{\em Phys.Rev.}
  {\bfseries D85} (2012) 104030},
\href{http://arxiv.org/abs/1109.3611}{{\ttfamily arXiv:1109.3611 [gr-qc]}}.

\bibitem{Hinderer:2013uwa}
T.~Hinderer, A.~Buonanno, A.~H. Mrou\'e, D.~A. Hemberger, G.~Lovelace, H.~P.
  Pfeiffer, L.~E. Kidder, M.~A. Scheel, B.~Szilagy, N.~W. Taylor, and S.~A.
  Teukolsky, ``{Periastron Advance in Spinning Black Hole Binaries: Comparing
  Effective-One-Body and Numerical Relativity},''
\href{http://arxiv.org/abs/1309.0544}{{\ttfamily arXiv:1309.0544 [gr-qc]}}.

\bibitem{Hinder:2013oqa}
{\bfseries The NRAR Collaboration, Perimeter Institute for Theoretical Physics}
  Collaboration, I.~Hinder {\em et~al.}, ``{Error-analysis and comparison to
  analytical models of numerical waveforms produced by the NRAR
  Collaboration},'' {\em Classical and Quantum Gravity} {\bfseries 31} no.~2,
  (2014) 025012, \href{http://arxiv.org/abs/1307.5307}{{\ttfamily
  arXiv:1307.5307 [gr-qc]}}.
\url{http://stacks.iop.org/0264-9381/31/i=2/a=025012}.

\bibitem{Taylor:2013zia}
N.~W. Taylor, M.~Boyle, C.~Reisswig, M.~A. Scheel, T.~Chu, L.~E. Kidder, and
  B.~Szil{\' a}gyi, ``{Comparing Gravitational Waveform Extrapolation to
  Cauchy-Characteristic Extraction in Binary Black Hole Simulations},''
  \href{http://dx.doi.org/10.1103/PhysRevD.88.124010}{{\em Phys. Rev. D}
  {\bfseries 88} (Dec, 2013) 124010},
  \href{http://arxiv.org/abs/1309.3605}{{\ttfamily arXiv:1309.3605 [gr-qc]}}.
  \url{http://link.aps.org/doi/10.1103/PhysRevD.88.124010}.

\bibitem{Reisswig:2012ka}
C.~Reisswig, N.~T. Bishop, and D.~Pollney, ``{General relativistic null-cone
  evolutions with a high-order scheme},'' {\em {Gen. Rel. Grav.}} {\bfseries 45
  (5)} (2013) 1069--1094,
\href{http://arxiv.org/abs/1208.3891}{{\ttfamily arXiv:1208.3891 [gr-qc]}}.

\bibitem{Babiuc:2010ze}
M.~C. Babiuc, B.~Szil\'agyi, J.~Winicour, and Y.~Zlochower, ``A characteristic
  extraction tool for gravitational waveforms,''
  \href{http://dx.doi.org/10.1103/PhysRevD.84.044057}{{\em Phys.\ Rev.\ D}
  {\bfseries 84} (Aug, 2011) 044057},
  \href{http://arxiv.org/abs/1011.4223}{{\ttfamily arXiv:1011.4223 [gr-qc]}}.
  \url{http://link.aps.org/doi/10.1103/PhysRevD.84.044057}.

\bibitem{Oliveira:2011}
{Oliveira, H. P.} and {Rodrigues, E. L.}, ``Numerical evolution of axisymmetric
  vacuum spacetimes: a code based on the galerkin method,'' {\em Class.\
  Quantum Grav.} {\bfseries 28} (2011) 235011,
  \href{http://arxiv.org/abs/0809.2837}{{\ttfamily arXiv:0809.2837 [gr-qc]}}.

\bibitem{Bishop:1998uk}
{Bishop, Nigel T.}, {Isaacson, R.}, {Gomez, R.}, {Lehner, L.}, {Szilagyi, B.},
  and {Winicour, J.}, ``{Cauchy-characteristic matching},''
\href{http://arxiv.org/abs/gr-qc/9801070}{{\ttfamily arXiv:gr-qc/9801070}}.

\bibitem{Szilagyi:2001fy}
B.~Szilagyi, B.~G. Schmidt, and J.~Winicour, ``{Boundary conditions in
  linearized harmonic gravity},''
  \href{http://dx.doi.org/10.1103/PhysRevD.65.064015}{{\em Phys.Rev.}
  {\bfseries D65} (2002) 064015},
\href{http://arxiv.org/abs/gr-qc/0106026}{{\ttfamily arXiv:gr-qc/0106026
  [gr-qc]}}.

\bibitem{Sachs1962}
R.~K. Sachs, ``Gravitational waves in general relativity. {VIII}. waves in
  asymptotically flat space-time,'' {\em Proc. R. Soc. Lond. A} {\bfseries 270}
  no.~1340, (October, 1962) 103--126.
  \url{http://www.jstor.org/stable/2416200}.

\bibitem{TamburinoWinicour1966}
L.~A. Tamburino and J.~H. Winicour, ``Gravitational fields in finite and
  conformal {B}ondi frames,'' {\em Phys. Rev.} {\bfseries 150} (1966)
  1039--1053. \url{http://link.aps.org/doi/10.1103/PhysRev.150.1039}.

\bibitem{hpgn}
N.~T. Bishop, R.~Gomez, L.~Lehner, M.~Maharaj, and J.~Winicour, ``High-powered
  gravitational news,'' {\em Phys. Rev.} {\bfseries D56} (1997) 6298--6309,
  \href{http://arxiv.org/abs/qr-qc/9708065}{{\ttfamily arXiv:qr-qc/9708065}}.

\bibitem{Loffler:2011ay}
F.~L{\" o}ffler, J.~Faber, E.~Bentivegna, T.~Bode, P.~Diener, {\em et~al.},
  ``{The Einstein Toolkit: A Community Computational Infrastructure for
  Relativistic Astrophysics},''
  \href{http://dx.doi.org/10.1088/0264-9381/29/11/115001}{{\em
  Class.Quant.Grav.} {\bfseries 29} (2012) 115001},
\href{http://arxiv.org/abs/1111.3344}{{\ttfamily arXiv:1111.3344 [gr-qc]}}.

\bibitem{Gomez:1997}
R.~G\'{o}mez, L.~Lehner, P.~Papadopoulos, and J.~Winicour, ``The eth formalism
  in numerical relativity,'' {\em Class.\ Quantum Grav.} (01, 1997) ,
  \href{http://arxiv.org/abs/gr-qc/9702002}{{\ttfamily arXiv:gr-qc/9702002}}.

\bibitem{Magnus1954}
W.~Magnus, ``On the exponential solution of differential equations for a linear
  operator,'' {\em Commun. Pure Appl. Math.} {\bfseries 7} (1954) 649--673.

\bibitem{MagnusReview2009}
{Blanes, S}, {Casas, F}, {Oteo, J A}, and {Ros, J}, ``The magnus expansion and
  some of its applications,'' {\em Phys. Rep.} {\bfseries 470 (5-6)} (January,
  2009) 151--238, \href{http://arxiv.org/abs/0810.5488}{{\ttfamily
  arXiv:0810.5488 [math-ph]}}.

\bibitem{Boyd1989}
J.~P. Boyd, {\em Chebyshev and Fourier Spectral Methods}.
\newblock Springer-Verlag, Berlin, 1989.

\bibitem{spherepack-home-page}
J.~C. Adams and P.~N. Swarztrauber, ``Spherepack 3.0.''.
  \url{http://www.scd.ucar.edu/css/software/spherepack}.

\bibitem{Huffenberger:2010}
{Huffenberger, Kevin M} and {Wandelt, Benjamin D}, ``Fast and exact spin-s
  spherical harmonic transforms,'' {\em Astro. J. Sup. Ser.} {\bfseries 189}
  (2010) 255--260, \href{http://arxiv.org/abs/1007.3514}{{\ttfamily
  arXiv:1007.3514 [astro-ph.IM]}}.

\bibitem{Babiuc2005}
M.~Babiuc, B.~Szil{\' a}gyi, I.~Hawke, and Y.~Zlochower, ``{Gravitational wave
  extraction based on Cauchy-characteristic extraction and characteristic
  evolution},'' {\em Class.\ Quantum Grav.} {\bfseries 22} no.~23, (2005)
  5089--5107, \href{http://arxiv.org/abs/gr-qc/0501008}{{\ttfamily
  arXiv:gr-qc/0501008}}. \url{http://stacks.iop.org/0264-9381/22/5089}.

\end{thebibliography}\endgroup

\appendix

\section{Regularizing Divergent Equations Using Integration by Parts}
This appendix describes the process of future null infinity regularization using integration by parts. A Laurent series is an expansion about a pole of (in this case) finite order. It is the logical extension of a Taylor series to functions that diverge in well defined ways.

We wish to radially integrate the following equation.

\begin{equation} 
(r^2 Q)_{,\rho} = A + \frac{B}{1-\rho} + \frac{C}{(1-\rho)^2} + \frac{D}{(1-\rho)^3}.
\end{equation}

Note that \(r = R \rho/(1-\rho)\), so both sides are infinite at \(\rho=1\).

The most divergent term is the \(D\) term - we integrate by parts.

\begin{equation}
\int \frac{D}{(1-\rho)^3} d\rho = \frac{D/2}{(1-\rho)^2} - \int \frac{D'/2}{(1-\rho)^2} d\rho\;.
\end{equation}

In practise, limits of integration are \(\rho \in [1/2,1]\) where \(\rho=1/2\) is the worldtube inner boundary of the domain.

This term is now an integral of the same order as the term in \(C\), as well as a term external to the integral that is of the same order. This is cancelled out through division by \(r^2 = R^2 \rho^2/(1-\rho)^2\). In this way, all the terms are regularized, ie. finite throughout the domain. We include the \(D'/2\) term in the \(C\) integral, and so on.

\begin{align}
\int \frac{-C+D'/2}{(1-\rho)^2} d\rho =& \int \frac{\partial}{\partial \rho} \left(\frac{1}{1-\rho}\right) (-C+D'/2) d\rho \nonumber\\
=& \frac{-C+D'/2}{1-\rho} - \int \frac{-C'+D''/2}{1-\rho} d\rho\;,
\end{align}
\begin{align}
\int &\frac{B-C'+D''/2}{1-\rho} d\rho = \nonumber\\
&-(B-C'+D''/2) \log(1-\rho) + \int (B'-C''+D'''/2)\log(1-\rho) d\rho\;,
\end{align}
\begin{align}
\int (B'-&C''+D'''/2)\log(1-\rho) d\rho = \nonumber\\
&-(B'-C''+D'''/2)(\rho+(1-\rho)\log(1-\rho)) \nonumber\\
&+ \int (B''-C'''+D''''/2)(\rho+(1-\rho)\log(1-\rho)) d\rho\;.
\end{align}

Crucially, the integral term is now bounded in the domain. This means it can be computed numerically. Computationally, the limits of integration are enforced by subtracting the value of the function on the inner boundary. \(Q_0\) is the boundary value of \(Q\). Combining terms, equation~3.11 can be expressed:

\begin{align}
Q = &Q_0 + \frac{D}{2 \rho^2 R^2} - (-C+D'/2)\frac{1-\rho}{\rho^2 R^2} - (B - C' + D''/2) \frac{(1-\rho)^2 \mathrm{log}(1-\rho)}{\rho^2 R^2} \nonumber\\
- &(B' - C'' + D'''/2)\frac{(\rho+(1-\rho)\mathrm{log}(1-\rho))(1-\rho)^2}{\rho^2 R^2} \nonumber\\
+ &\frac{(1-\rho)^2}{\rho^2 R^2} \int\; A + (B'' -C''' + D''''/2)(\rho+(1-\rho)\mathrm{log}(1-\rho))\;. 
\end{align}

\section{Inner Boundary Algorithm Formalism}
The boundary algorithm formalism is drawn from Bishop {\it et al.}\cite{Bishop:1998uk}. It begins with metric quantities on the worldtube forming the boundary between the space-like foliated numerical GR simulation and the null-foliated CCE domain. It calculates several intermediate helper quantities to simplify the computational complexity. Finally, it produces boundary values for each of the hypersurface or Bondi metric quantities. Note that all quantities in this section refer to inner boundary values only.

\subsection{Initial Metric Quantities}
The Metric Quantities are extracted directly from the simulation. They are the covariant 3-metric \(g_{ij}\), the contravariant 3-metric \(g^{ij} = \left(g_{ij}\right)^{-1}\), the co- and contravariant 3-metric derivatives \(g_{ij},_\alpha\) and \(g^{ij},_\alpha\) (calculated using \(g^{ij},_\gamma = - g^{ik} g^{jl} g_{kl},_\gamma\)), and their 4-metric counterparts. The time components are calculated from the lapse \(\alpha = \sqrt{g_{i0}\beta^i - g_{00}}\) and shift \(\beta^i = g^{ij} g_{j0}\) as follows.
\begin{align}
g_{k0} =& g_{0k} = g_{ki}\beta^i\;,\\
g_{00} =& g_{i0}\beta^i - \alpha^2 = g_{ik}\beta^i \beta^k - \alpha^2\;\\
g_{k0},_\gamma =& g_{0k},_\gamma = (g_{ki} \beta^i),_\gamma = g_{ki},_\gamma \beta^i + g_{ki}\beta^i,_\gamma\;,\\
g_{00},_\gamma =& g_{i0},_\gamma \beta^i + g_{i0}\beta^i,_\gamma - 2\alpha \alpha,_\gamma\;,\\
g^{\alpha \beta},_\gamma =& - g^{\alpha \delta} g^{\beta \epsilon} g_{\delta \epsilon},_\gamma\;.
\end{align}

\subsection{Intermediate Quantities}
Additional spherical derivates are calculated by evaluating the angular derivatives of the spherical harmonic expansions of the quantities on the worldtube.

\subsubsection{Jacobians}
To perform coordinate transformations, Jacobians and their derivatives were derived. They included the spherical-to-Cartesian \(\Lambda^{(r,A)}_i\) and its derivative \(\Lambda^{(r,A)}_i,_\alpha\), the Cartesian-to-spherical \(\Lambda^i_{(r,A)}\) and its derivative \(\Lambda^i_{(r,A)},_\alpha\).

\subsubsection{Null Generator}
We calculate the unit normal vectors to the time slice \((n^\gamma)\) and the sphere \((s^\gamma)\) to calculate the null generator \((l^\gamma)\). 
\begin{align}
n^\gamma =& \left( 1/\alpha, -\beta^i/\alpha \right)\;,\\
s^\gamma =& (0,\frac{g^{ij} x_j}{\sqrt{g^{ij} x_i x_j}})\;,\\
l^\gamma =& \frac{n^\gamma + s^\gamma}{\alpha - \beta^i s^j g_{ij}}\;.\\
\end{align}

Their time derivatives are
\begin{align}
n^\alpha,_t =& \frac{1}{\alpha^2}(-\alpha,_t,\alpha,_t \beta^i - \alpha \beta^i,_t)\;,\\
s^i,_t =& (-g^{im} + s^i s^m/2)g_{mn},_t s^n\;,\\
s^\alpha,_t =& \left(0,s^i,_t\right)\;,\\
l^\alpha,_t =& \frac{n^\alpha,_t + s^\alpha,_t - l^\alpha \alpha,_t + l^\alpha(g_{ij},_t \beta^i s^j + g_{ij} \beta^i,_t s^j + g_{ij} \beta^i s^j,_t)}{\alpha - g_{ij} \beta^i s^j}\;.
\end{align}

\subsubsection{(Affine) Spherical Metric Quantities}
Dramatic simplification is possible by calculating a number of auxiliary metric terms in a spherical coordinate system. Here, \(,_\lambda\) denotes a derivative in the null direction, whereas \(A\), \(B\), \(C\) etc sub- and superscripts denote indexes across the spherical components of the coordinate system, i.e. \(\theta\) and \(\phi\).

\begin{align}
g_{\alpha \beta},_\lambda =& l^\gamma g_{\alpha \beta},_\gamma\;,\\
\eta_{AB} =& \Lambda^i_A \Lambda^j_B g_{ij}\;,\\
\eta_{AB},_\lambda =& \Lambda^k_A \Lambda^l_B g_{kl},_\lambda + (l^\mu,_A \Lambda^l_B + l^\mu,_B \Lambda^l_A) g_{\mu l}\;,\\
\eta_{AB},_t =& \Lambda^i_A \Lambda^j_B g_{ij},_t\;,\\
\eta_{uA},_\lambda =& l^t,_A g_{tt} + \Lambda^k_A g_{tk},_\lambda + l^k,_A g_{tk} + \Lambda^k_A l^\mu,_t g_{\mu k}\;.
\end{align}

The contravariant quantities are similarly defined.
\begin{align}
\eta^{AB}\eta_{BC} =& \delta^A_C\;,\\
\eta^{A \lambda} =& \eta^{AB} \Lambda^k_B g_{tk}\;,\\
\eta^{\lambda \lambda} =& -g_{tt} + \eta^{A \lambda} \Lambda^k_A g_{tk} \;,\\
\eta^{AB},_\lambda =& -\eta^{AC} \eta^{BD} \eta_{CD},_\lambda\;,\\
\eta^{A\lambda},_\lambda =& \eta^{AB} \eta_{uA},_\lambda - \eta^{AB} \eta^{C \lambda} \eta_{CB},_\lambda\;.
\end{align}

\subsubsection{Dyad Quantities}
The dyad allows construction of spin-weighted scalars on the sphere. Given \(r = \sqrt{x^2+y^2+z^2},\) 

\begin{align}
q^i =& \frac{1}{r\sqrt{x^2 + y^2}}\left( -x z + i y r, -y z - i x r, x^2+y^2 \right)\;,\\
q_A =& \Lambda^i_A q^i/r\;,
\end{align}
since \(q_i = q^i\). Explicitly,
\begin{align}
q_A =& \left(-1,-i\frac{r_c}{r}\right)\;,\\
q^A =& \left(-1,-i\frac{r}{r_c}\right)\;,
\end{align}
where \(r_c = \sqrt{x^2+y^2},\) and its derivatives are defined as follows.

\begin{align}
q^i_{,t} = \frac{1}{r^3 r_c^3}\left(\right.
&z((x^4-y^2(y^2+z^2))x_{,t} + x y ( 2r_c^2+z^2)y_{,t}) - x r_c^4 z_{,t} 
+ i r^3(x(-y x_{,t} + x y_{,t})),\nonumber\\
&z(x y (2 r_c^2+z^2)x_{,t}-(x^4-y^4+x^2 z^2)y_{,t})-y r_c^4 z_{,t} 
+ i r^3(y(-y x_{,t} + x y_{,t})),\nonumber\\
&\left.r_c^2(z(z(x x_{,t}+y y_{,t})-r_c^2 z_{,t}))\right)\;,
\end{align}

\begin{align}
q^i_{,x} =& \frac{1}{r^3 r_c^3}\left(z(x^4-y^2(y^2+z^2)) -i r^3 x y,
                                    x y z(2r_c^2+z^2) - i r^3 y^2,
                                    x z^2 r_c^2 \right)\;,\\
q^i_{,y} =& \frac{1}{r^3 r_c^3}\left(x y z(2r_c^2+z^2) + i x^2 r^3,
                                    -z(x^4-y^4+x^2 z^2) + i x y r^3,
                                    y z^2 r_c^2\right)\;,\\
q^i_{,z} =& \frac{r_c}{R^3}\left(-x,-y,-z\right)\;,
\end{align}
Then \(q^i,_\lambda = l^\alpha q^i,_\alpha\). Note that the \(q^i\) are not necessarily constant \((q^i_{,t} \neq 0)\) as the properties of the worldtube are time-dependent. This approach differs from the Pitt null code.

\subsubsection{The Bondi r}
The Bondi radius \(r_b\) is an areal radius, not a coordinate radius. The value of the Bondi radius \(r_b\) at every point on the worldtube provides the compactification parameter. Given \(|q_{AB}| = \sin^2\theta = \frac{x^2+y^2}{x^2+y^2+z^2},\)
\begin{align}
r_b =& \left(\frac{|\eta_{AB}|}{|q_{AB}|}\right)^\frac{1}{4}\;,\\
r_b,_\lambda =& \frac{r_b}{4} \eta^{AB} \eta_{AB},_\lambda\;,\\
r_b,_t =& \frac{r_b}{4} \eta^{AB} \eta_{AB},_t\;.
\end{align}

In contrast, derivatives of the coordinate radius \(r\) are 
\begin{align}
r,_\lambda =& l^\alpha r,_\alpha\;,\\
r,_t =& \frac{x x,_t + y y,_t + z z,_t}{r}\;,\\
r,_\alpha =& \frac{1}{r}\left(x^i x_i,_t,x^i\right)\;.
\end{align}
In the present algorithm the cartesian extraction radius \(r\) is held constant, but derivatives are included for the general case of a variable extraction coordinate radius.

\subsubsection{Derivatives of Hypersurface Quantities}
Calculation of \(Q\) and \(\Phi = J,_u\) require the null-derivatives of several other hypersurface quantities.
\begin{align}
\beta,_\lambda =& \frac{r_b}{8 r_b,_\lambda} \left(J,_\lambda \bar{J},_\lambda - \frac{1}{1+J \bar{J}}\mathbb{R}(\bar{J} J,_\lambda)^2 \right)\;,\\
J,_\lambda =& 2 J \left(\frac{r,_\lambda}{r} -\frac{r_b,_\lambda}{r_b}\right) + \frac{r^2}{r_b^2} (g_{ij} q^i,_\lambda q^j + \frac{1}{2} g_{ij},_\lambda q^i q^j)\;,\\
U,_\lambda =& 2 \beta,_\lambda U + 2\beta,_\lambda \eta^{A\lambda}q_A - \eta^{A\lambda},_\lambda q_A - r_b,_{\lambda B} \eta^{AB} q_A + r_b,_B \eta^{AB},_\lambda q_A\;.
\end{align}

\subsection{Hypersurface Quantities}
The hypersurface quanties are \(J\), \(K\), \(\beta\), \(U\), \(Q\) and \(W\), as well as the time derivative \(J,_u\).

\begin{align}
J =& \frac{r^2}{2 r_b^2} g_{ij} q^i q^j\;,\\
\Phi =& J,_u=2 J \left(\frac{r,_t}{r} -\frac{r_b,_t}{r_b}\right) + \frac{r^2}{r_b^2} (g_{ij} q^i,_t q^j + \frac{1}{2} g_{ij},_t q^i q^j)\;,\\
K =& \sqrt{1+J\bar{J}}\;,\\
\beta =& -\frac{1}{2} log(r_b,_\lambda)\;,\\
U =& -(\eta^{\lambda A} + \frac{r_b,_B}{r_b,_\lambda} \eta^{AB})q_A\;,\\
Q =& r_b^2(J \bar{U},_\lambda + \sqrt{1+J \bar{J}} U,_\lambda) \;,\\
W =& \frac{r_b,_\lambda \eta^{\lambda \lambda}}{r_b} - \frac{2 r_b,_u}{r_b} - \frac{1}{r_b} + \frac{2 r_b,_A \eta^{A \lambda}}{r_b} + \frac{r_b,_A r_b,_B \eta^{AB}}{r_b,_\lambda r_b} \;.
\end{align}

\subsection{Bondi Metric Reconstruction}
\(J\), \(K\), \(\beta\), \(U\), \(Q\) and \(W\) can be combined to reconstruct the Bondi metric, with standard coordinate 4-vector ordering \((u,r,\theta,\phi)\)\cite{hpgn}.
\begin{equation}
\eta^{\alpha \beta} = 
\begin{pmatrix}
0 & -e^{-2\beta} & 0 & 0 \\
-e^{-2\beta} & (r W + 1)e^{-2\beta} & \frac{1}{2}(U+\bar{U})e^{-2\beta} & -\frac{i r}{2 r_c}(U-\bar{U})e^{-2\beta} \\
0 & \frac{1}{2}(U+\bar{U})e^{-2\beta} & \frac{1}{2 r^2}(2K-J-\bar{J}) & \frac{i}{2 r r_c}(J-\bar{J}) \\
0 & -\frac{i r}{2 r_c}(U-\bar{U})e^{-2\beta} & \frac{i}{2 r r_c}(J-\bar{J}) & \frac{1}{2 r_c^2} (2K + J + \bar{J}) 
\end{pmatrix}
\;,
\end{equation}

\begin{equation}
\eta_{\alpha \beta} =
\begin{pmatrix}
  -(r W + 1)e^{2\beta}  & & \frac{r^2}{2}\left((J+K)\bar{U} \right.  & \frac{i r r_c}{2} \left(\bar{J} U - J\bar{U} \right. \\
  + \frac{r^2}{2}\left(2 K U \bar{U} + J\bar{U}^2 + \bar{J} U^2\right) & -e^{2\beta} & + \left. (\bar{J}+K)U \right) & + \left. K (\bar{U}-U)\right) \\
  & & & \\
  -e^{2\beta} & 0 & 0 & 0 \\
  \frac{r^2}{2}\left((J+K)\bar{U} + (\bar{J}+K)U \right) & 0 & \frac{r^2}{2}(J + \bar{J} + 2K) & \frac{i r r_c}{2}(\bar{J}-J) \\
  \frac{i r r_c}{2} \left(\bar{J} U - J\bar{U} +K (\bar{U}-U)\right) & 0 & \frac{i r r_c}{2}(\bar{J}-J) & -\frac{r_c^2}{2}(J + \bar{J} - 2K)
\end{pmatrix}
\;.
\end{equation}

\section{\(\eth\) Operator and Spin-Weighted Spherical Harmonics}
The hypersurface equations are written making use of the \(\eth\) formalism\cite{Gomez:1997}, which simplifies equations with spherical symmetry. In our implementation, the we use a basis capable of multiple values on the poles, so only a single patch is necessary to cover the entire sphere.

Rank 2 tensors on the sphere can be broken down and expressed as a sum of spin-weighted spherical harmonics when contracted with an appropriate dyad \(q^A\). Following \cite{Gomez:1997}(Eqn.~8), a generic tensor can be written as the sum of respectively a symmetric trace free part, a trace part, and an antisymmetric part:
\begin{equation}
w_{AB} = t_{AB} + \frac{p}{2} q_{AB} + i \frac{u}{4}\left(q_A \bar{q_B} - \bar{q_A} q_B\right)\;.
\end{equation}
\(p = q^{CD} w_{CD}\) and \(u = i(q^C \bar{q}^D - \bar{q}^C q^D) w_{CD}/2\) are both real scalar fields with spin-weight 0. Writing \(t = t_{AB} q^A q^B\) yields a complex scalar field with spin-weight 2. Together, these three scalars \(p\), \(u\) and \(t\) completely specify the tensor field independent of choice of basis.

\(\eth\) is a spherical derivative operator on spin-weighted spherical harmonics. In spherical coordinates
\begin{align}
\eth\eta = -(\sin^s \theta) \left(\frac{\partial}{\partial \theta} + \frac{i}{\sin\theta}\frac{\partial}{\partial \phi}\right)(\sin^{-s}\theta \, \eta)\;,\\
\bar{\eth}\eta = -(\sin^{-s} \theta) \left(\frac{\partial}{\partial \theta} - \frac{i}{\sin\theta}\frac{\partial}{\partial \phi}\right)(\sin^s\theta \, \eta)\;.
\end{align}
This can be thought of as a contraction of the dyad with the spherical derivatives operator.

In terms of spin-weighted spherical harmonics
\begin{align}
\eth\left(_sY_{lm}\right) = + \sqrt{(l-s)(l+s+1)}_{s+1}Y_{lm}\;,\\
\bar{\eth}\left(_sY_{lm}\right) = - \sqrt{(l+s)(l-s+1)}_{s+1}Y_{lm}\;.
\end{align}

Used in combination, these definitions allow the \(\eth\) formalism to be used in the spectral domain. We used the package Spinsfast\cite{Huffenberger:2010} to implement spin-weighted spherical harmonics. 

\(\eth\) assumes a spherical domain. However, in general the compactification parameter \(R=r_{b|\Gamma}\) is not constant on the sphere. The corrected operation is

\begin{equation}
\eth \eta =  \tilde \eth \eta - \eta,_\rho \frac{\rho(1-\rho)}{R} \tilde \eth R \;,
\end{equation}

where \(\tilde \eth\) denotes the naive operator in the affine coordinate system.

This correction is the same that operates at the end of the calculation of \(J,_u = \Phi\) in Appendix D, only in reverse\cite{Babiuc2005}.

\section{All Evolution Algorithm Terms}
These are largely drawn from Bishop {\it et al.}\cite{hpgn}, with some re-arrangement of terms to ensure internal consistency with the Magnus expansion formalism, and the compactification transformation \(\rho = r/(R+r)\). Additionally, terms are grouped by their order in the Laurent expansion, where relevant.

\subsection{\(\beta\) terms}
\begin{equation}
\beta_{,r} = N_\beta\;,
\end{equation}

\begin{equation}
N_\beta = \frac{\rho(1-\rho)^3}{8 R}(J,_\rho \bar{J},_\rho - K^2,_\rho)\;.
\end{equation}

\subsection{Q terms}
\begin{equation}
(r^2 Q)_{,\rho} = \frac{Q_C+Q_{CNL}}{(1-\rho)^2} + \frac{Q_D}{(1-\rho)^3}\;.
\end{equation}

\begin{align}
Q_C =& -R^2 \rho^2 \bar{\eth}J,_\rho - R^2 \rho^2 \eth K,_\rho + 2 R^2\rho^2 \eth \beta,_\rho\;,\\
Q_{CNL} =& R^2 \rho^2 (1-K)(\eth K,_\rho + \bar{\eth} J,_\rho) + \bar{J}\eth J,_\rho + J,_\rho \bar{J},_\rho + J \bar{\eth}K,_\rho \nonumber\\
&+ K,_\rho \bar{\eth}J - J,_\rho \bar{\eth}K + \frac{\eth \bar{J}(J,_\rho - J^2 \bar{J},_\rho) + \eth J(\bar{J},_\rho - \bar{J}^2 J,_\rho)}{2 K^2}\;,\\
Q_D =& -4 R^2 \rho \eth\beta\;.
\end{align}

\subsection{U terms}
\begin{equation}
U,_\rho = U_A + U_{ANL}\;.
\end{equation}

\begin{align}
U_A =& \frac{e^{2\beta}}{R \rho^2}Q\;,\\
U_{ANL} =& \frac{e^{2\beta}}{R \rho^2} (K Q-Q-J\bar{Q})\;.
\end{align}

\subsection{W terms}
\begin{equation}
(r^2 W),_\rho = \frac{W_C+W_{CNL}}{(1-\rho)^2} + \frac{W_D}{(1-\rho)^3}\;.
\end{equation}

\begin{align}
W_C =& \mathcal{R} - R^2 e^{2\beta}(\eth \beta \bar{\eth}\beta+\eth \bar{\eth}\beta) + \frac{R^2 \rho^2}{4} (\eth \bar{U},_\rho + \bar{\eth}U,_\rho)\;, \;\mathrm{where} \\
\mathcal{R} =& \frac{R e^{2 \beta}}{2}\left( 2K - \eth \bar{\eth}K + \frac{1}{2}(\bar{\eth}^2J + \eth^2 \bar{J}) + \frac{1}{4K} (\bar{\eth}\bar{J} \eth J - \bar{\eth}J \eth \bar{J}) \right)\;, \\
W_{CNL} =& R e^{2\beta} \left( (1-K)(\eth\bar{\eth}\beta +\eth\beta\bar{\eth}\beta) + \frac{1}{2}(J \bar{\eth}\beta^2+\bar{J}\eth\beta^2) \right. \nonumber\\
& \left. - \frac{1}{2}(\eth\beta(\bar{\eth}K - \eth\bar{J}) + \bar{\eth}\beta(\eth K - \bar{\eth}J)) + \frac{1}{2}(J \bar{\eth}^2 \beta + \bar{J} \eth^2 \beta)\right) \nonumber\\
& - e^{2\beta}\frac{R^3 \rho^4}{8}\left(2 K U,_\rho \bar{U},_\rho + J \bar{U}^2,_\rho + \bar{J} U^2,_\rho\right)\;,\\
W_D =& R^2 \rho (\eth \bar{U} + \bar{\eth} U)\;.
\end{align}

\subsection{\(J,_u = \Phi\) terms}
\begin{equation}
(r \Phi),_\rho - (r J) (\Phi \bar{\Gamma} + \bar{\Phi} \Gamma) = \Phi_A + \Phi_{ANL} + \frac{\Phi_B + \Phi_{BNL}}{1-\rho} + \frac{\Phi_C+\Phi_{CNL}}{(1-\rho)^2}\;.
\end{equation}
\begin{equation}
\Gamma = \left(J,_\rho - J \frac{K,_\rho}{K}\right)\;.
\end{equation}

\begin{align}
\Phi_A =& (1-\rho) J,_\rho + \frac{1}{2} R \rho^2 W,_\rho J,_\rho + \frac{\rho}{2}(1-\rho+R\rho W)J,_{\rho\rho}\;,\\
\Phi_B =& (\frac{3}{2} R \rho - R \rho^2)W J,_\rho - \frac{1}{2} R \rho \eth U,_\rho + \frac{e^{2\beta}}{\rho}(\eth^2\beta + \eth\beta^2)\;,\\
\Phi_C =& -R \eth U\;,\\
\Phi_{ANL} =& -4 J \beta,_\rho\;,\\
\Phi_{BNL} =& N_{1B} + N_{2B} + N_{3B} + N_{4B} + N_{5B} + N_{6B} + N_{7B} \nonumber\\
&+ P_{1B} + P_{2B} + P_{3B} + P_{4B}\;,\\
\Phi_{CNL} =& N_{2C} + N_{3C} + P_{3C}\;,
\end{align}

\(N\) terms are as in Bishop {\it et al.}\cite{hpgn}, with a prefactor \(\frac{R}{2(1-\rho)^2}\) and the usual compactification transformation.

\begin{align}
N_{1B} =& -\frac{e^{2\beta}}{2\rho}\left(K(\eth J \bar{\eth}\beta + 2\eth K \eth \beta - \bar{\eth}J \eth \beta) \right. \nonumber\\
&\left.+ J(\bar{\eth}J \bar{\eth}\beta - 2\eth K \bar{\eth} \beta) - \bar{J} \eth J \eth \beta \right)\;,\\
N_{2B} =& -\frac{R \rho}{4} \left( \eth J \bar{U},_\rho + \bar{\eth} J U,_\rho\right)\;,\\
N_{2C} =& -\frac{R}{2}\left(\eth J \bar{U} + \bar{\eth} J U\right)\;,\\
N_{3B} =& \frac{R \rho}{2}\left( (1-K)\eth U,_\rho - J \eth \bar{U},_\rho\right)\;,\\
N_{3C} =& R(1-K) \eth U - R J \eth \bar{U}\;,\\
N_{4B} =& \frac{e^{-2 \beta}}{4} R^2 \rho^3 \left(K^2 U^2,_\rho + 2 J K U,_\rho \bar{U},_\rho + J^2 \bar{U}^2,_\rho\right)\;,\\
N_{5B} =& -\frac{R \rho}{4} J,_\rho(\eth \bar{U} + \bar{\eth} U)\;,\\
N_{6B} =& \frac{R \rho}{2} \left(\frac{1}{2}(\bar{U} \eth J + U \bar{\eth}J)(J \bar{J},_\rho - \bar{J} J,_\rho) + \bar{U} \bar{\eth} J (J K,_\rho - K J,_\rho) \right. \nonumber\\
&\left.- \bar{U}(\eth J,_\rho - 2 K \eth K J,_\rho + 2 J \eth K K,_\rho) - U(\bar{\eth} J,_\rho - K \eth \bar{J} J,_\rho + J \eth \bar{J} K,_\rho) \right)\;,\\
N_{7B} =& \frac{R \rho}{2} (K J,_\rho - J K,_\rho) \left(\bar{U} (\bar{\eth} J - \eth K) + U (\bar{\eth} K - \eth \bar{J}) \right. \nonumber\\
&\left.+ K (\bar{\eth}U - \eth \bar{U}) + (J \bar{\eth} \bar{U} - \bar{J} \eth U) \right)\;.
\end{align}

\(P\) terms are as in Bishop {\it et al.}\cite{hpgn} except for the terms in \(J,_u\), which have been moved to the left hand side of the equation. They have a prefactor \(\frac{J}{2\rho(1-\rho)}\) and the usual compactification transformation.
\begin{align}
P_{1A} =& -4 J \beta,_\rho\;,\\
P_{1B} =& -4 R \rho J \beta,_\rho W\;,\\
P_{2B} =& \frac{e^{2\beta}}{2 \rho} J \left( -2 K (\eth \bar{\eth} \beta + \bar{\eth} \beta \eth \beta) - (\bar{\eth} \beta \eth K + \eth \beta \bar{\eth} K) + J( \bar{\eth}^2 \beta + \bar{\eth} \beta^2) \right. \nonumber\\
&\left. + \bar{J}(\eth^2\beta + \eth \beta^2) + (\bar{\eth} J \bar{\eth}\beta + \eth \bar{J} \eth \beta)\right)\;,\\
P_{3B} =& \frac{R \rho}{4} J (\bar{\eth} U,_\rho + \eth \bar{U},_\rho)\;,\\
P_{3C} =& \frac{R}{2} J (\bar{\eth} U + \eth \bar{U})\;,\\
P_{4B} =& -\frac{e^{-2 \beta}}{2}R^2 \rho^3 J \left(2 K U,_\rho \bar{U},_\rho + J \bar{U}^2,_\rho + \bar{J} U^2,_\rho \right)\;.
\end{align}

The non-spherical and non-constant inner boundary creates a discrepancy between the Bondi and affine coordinates. The corrective factor is given by 
\begin{equation}
f,_{\tilde u} = f,_u + f,_r r,_u = f,_u + \rho(1-\rho) \frac{R,_u}{R} f,_\rho\;,
\end{equation}
where once again \(,_{\tilde u}\) denotes a derivative performed in the affine coordinate system, whereas~\(,_u\) is the regular derivative in the Bondi coordinate system. Time steps are performed in the affine coordinate system, whereas the hypersurface equations are calculated and solved in the Bondi coordinate system.

This term must be added at the end of the calculation, in any instance where \(R,_u\) is nonzero.

\end{document}